\documentclass[12pt,preprint]{aastex}
 \usepackage{multirow}
 \usepackage{amsmath}
  \usepackage{epstopdf}
  \usepackage{booktabs}
  \usepackage{graphics}
  \usepackage{epsfig}
  \usepackage{subfigure}

\shorttitle{Black Holes In Young Stellar Clusters}
\shortauthors{Goswami et al.}
\begin{document}

\title{Black Holes In Young Stellar Clusters}

\author{Sanghamitra Goswami,\altaffilmark{1}  Paul Kiel,\altaffilmark{1,2}  Frederic A.\ Rasio\altaffilmark{1,3}}
\affil{$^{1}$ Department of Physics and Astronomy, Northwestern University, 
Evanston, IL 60208, USA}
\affil{$^{2}$ Monash Centre for Astrophysics, Monash University, Victoria 3800, Australia}
\affil{$^{3}$ Center for Interdisciplinary Exploration and Research in Astrophysics (CIERA), Northwestern University.}

\begin{abstract} 
  We present theoretical models for stellar black hole (BH) properties in young, massive star clusters. Using a Monte Carlo code for stellar dynamics, we model realistic star clusters with $N\simeq 5\times10^5$ stars and significant binary fractions (up to $50\%$) with self-consistent treatments of stellar dynamics and stellar evolution. We compute the formation rates and characteristic properties of single and binary BHs for various representative ages, cluster parameters, and metallicities. Because of dynamical interactions and supernova (SN) kicks, more single BHs end up retained in clusters compared to BHs in binaries. We also find that the ejection of BHs from a cluster is a strong function of initial density. In low-density clusters (where dynamical effects are negligible), it is mainly SN kicks that eject BHs from the cluster, whereas in high-density clusters (initial central density $\rho_c(0) \sim 10^5 \, M_\odot\, {\rm pc}^{-3} $ in our models) the BH ejection rate is enhanced significantly by dynamics. Dynamical interactions of binary systems in dense clusters also modify the orbital period and eccentricity distributions while also increasing the probability of a BH having a more massive companion. 
 \end{abstract}



\section{Introduction}

Young massive clusters (YMCs) and super star clusters (SSCs) are dense systems of young stars, which have received considerable interest over the past decade. Typically, YMCs are younger than $\sim100\,\rm Myr$, more massive than $\sim10^{4}\, M_{\odot}$ and have a density higher than $\sim10^{3}\, M_{\odot}/\rm pc^{3}$ \citep{Zwa10}. Observationally, YMCs are found in a variety of environments including the Milky Way Galaxy and the Local Group, starburst galaxies and interacting galaxies. SSCs ($\ge10^5\,M_{\odot}$) populate the most massive and luminous end in the continuum of YMCs. \citet{Elm97} and others have argued that the upper end of the luminosity function of globular clusters (GCs) is very similar to that observed for YMCs, suggesting that a universal formation mechanism might be responsible for the formation of star clusters in all environments and at all epochs.

 In addition to providing predictions for X-ray sources in YMCs and SSCs, the theoretical study of primordial BH populations in clusters can provide realistic initial conditions for investigating the long-term dynamical evolution of GCs. Mass segregation in GCs has often been studied previously using simple two-component models, where the BHs are treated as a tracer population of heavy objects in a background of equal-mass, lighter stars (e.g., \citet{Fre02,Bre13}). Mass segregation is a consequence of the tendency towards equipartition of energy among stars of different masses. As the cluster evolves, the more massive objects (here BH remnants) concentrate lower in the gravitational potential well, i.e., closer to the cluster center, whereas the lighter stars gain energy and are displaced outward. While previous studies have shed light on this dynamical process in the context of simplified models, a more realistic study of BHs in GCs clearly requires a treatment of the full stellar mass spectrum \citep{Mor13}.

Past theoretical work on BHs in clusters has also been done using direct $N$-body simulations; these are always limited to small-$N$ systems, however, in a range more appropriate for modeling {\em open clusters\/}.
\citet{Por00} investigated the formation of BH--BH binaries in star clusters through dynamical encounters, with a focus on predicting the merger rate for these binaries, which are key sources of gravitational waves detectable by the current generation of laser interferometers such as LIGO \citep{Aas13}. They found that, in their models with $N\sim 10^3$, BHs sink to the core within $\sim 10\,\rm Myr$. In the core these BHs acquire binary companions; the binaries harden quickly through superelastic encounters with other stars or BHs and are ultimately ejected from the cluster. They predicted that the ejected BH binaries have very short orbital periods and high eccentricities, and will coalesce within a few Gyr, contributing significantly to the potential LIGO detection rate.

\citet{Mac07} studied the effect of stellar mass BHs on the dynamical evolution and structural parameters of Magellanic Cloud clusters using larger $N$-body simulations with $N\sim10^5$. However, their simulations did not include primordial binaries and assumed simple limiting cases of natal kicks imparted to BHs. They argued that
both mass loss from early stellar evolution and longer-term heating by BH ejections and interactions can produce a core expansion with age, in agreement with observations of Magellanic Cloud clusters.
More recently, \citet{Map13} used direct $N$-body simulations with $N\sim 10^4$ to study the dynamics of stellar BHs in young star clusters with different metallicities, focusing on the implications for X-ray binaries. We will return to this study and compare it to our results for much larger clusters in Sections~2 and~4.

{\em Chandra\/} observations of starburst galaxies uncovered large numbers of bright point X-ray sources \citep{Fab01,Kaa01}. Optical and infrared observations of these sources often revealed massive young clusters or SSCs associated with them \citep{Zez02}. From previous theoretical studies and the spectral and temporal variability of these sources, it is now believed that many of these sources are bright X-ray binaries formed in star clusters \citep{Kaa01}. From observations of the Antennae it has been concluded that the luminosity ($L_{X}$) of these X-ray sources can be divided into $3$ classes \citep{Zezlum}. The lowest luminosity sources with $L_{X}<3\times10^{38}\,\rm  erg\,s^{-1}$ are thought to be SN remnants or neutron star binaries from their steep X-ray spectra. Power-law spectra of sources with luminosities exceeding the Eddington luminosity for neutron star binaries ($3\times10^{38}<L_{X}<10^{39}\,\rm ergs\,s^{-1}$) resemble those of Galactic BH X-ray binaries (XRBs). Out of the $18$ sources in this luminosity range in the Antennae, $10$ sources are thought to be associated with a young star cluster (age $<100\,\rm Myr$). If $L_{X}>10^{39}\,\rm ergs\,s^{-1}$, the sources are classified as Ultraluminous X-ray sources (ULXs), possibly associated with more massive stellar BHs \citet{Bel04}
or even intermediate-mass BHs (IMBHs). Out of the $49$ X-ray sources detected in the Antennae, 18 are ULXs.

\citet{Kaa04} found significant statistical association between X-ray point sources and young stellar populations in the three starburst galaxies (M82, NGC 1569, and NGC 5253) they studied: the X-ray point sources are at distances of $30-100\,\rm pc$ from the young star clusters they are associated with. However, there is an apparent lack of X-ray sources coincident with the clusters. \citet{Sep04} tried to explain this observation by arguing that a significant number of X-ray sources can be ejected from the parent cluster due to SN kicks. However, they used a simple population synthesis approach to study the effects of SN kicks on XRB populations, neglecting any cluster dynamics.

{\em Chandra\/} X-ray observations of the Galactic young cluster Westerlund 1 (Wd~1) have revealed a large population of X-ray sources \citep{Cla08}. Wd~1 has an age of $4-5\,\rm Myr$ and is estimated to have a high binary fraction \citep{Ski06,Cla11}. Observations of Wolf Rayet and OB stars in Wd~1 have suggested a binary fraction approaching unity \citep{Cro06} for massive stars (with initial masses $> 45\, M_{\odot}$). In spite of the rich X-ray point source population in Wd~1, there is a lack of  bright X-ray sources ($L_{X}>10^{38}\,\rm  erg\,s^{-1}$) in this cluster. This is in spite of the young cluster age, which would seem to imply the existence of many high-mass XRBs accreting from stellar winds of massive
main-sequence companions. Several possible reasons have been offered to explain this discrepancy \citep{Cla05,Neg05},
all involving combinations of binary stellar evolution and dynamics.

Clearly, a more detailed theoretical study of BHs in young star clusters is necessary to explain the statistics, spatial distributions, and luminosities of these X-ray sources.
Our goal in this paper is to model, in a self-consistent manner, the populations of BHs in YMCs and SSCs, taking into account both stellar dynamics and stellar evolution. This is only a first step: detailed modeling of X-ray sources is beyond the scope of this study, which will focus instead on understanding the effects of dynamics in large clusters  and the overall properties of all single and binary BHs, independent of whether they might be active X-ray sources or not.
The initial conditions and stellar evolution prescriptions in our work include a full global initial mass function, significant binary fractions, and realistic natal kick prescriptions for BHs based on our current understanding of core collapse SNe.

Our paper is organized as follows. In Section~2 we discuss in more detail and summarize some of the previous theoretical work addressing this problem. In Section~3 we provide a detailed description of our stellar dynamics code and our stellar evolution prescriptions. In Section~4 we present and discuss our results. Finally, summary and conclusions are presented in Section~5.

\section{ Previous Studies}

\subsection{BHs in Young Starburst Environments}

\citet[][thereafter B04]{Bel04} studied young populations of BHs as observed in starburst galaxies using a population synthesis approach. Their simulations did not explicitly take into account that the starburst was happening within a clustered environment. They investigated, for many representative models, the numbers of BH systems produced as well as their physical properties (e.g., binary period and BH mass distributions). They also studied, in great detail, the evolutionary channels leading to these different properties.
They found that, soon after (within $\sim 5\,\rm Myr$)
the initial starburst most BHs were single, with only $\sim10\%$ in binaries. Single BHs were also formed from binary progenitors, when, e.g., the binary disrupted following a SN explosion, or underwent a merger following Roche lobe overflow (RLOF) and dynamically unstable mass transfer. In the initial few Myr,
the most common BH binary systems were BH--MS
systems. After the cluster had evolved for a significant period of time ($> 20\,\rm Myr$)
the most common BH binary systems were BH--BH systems.
In B04, the overall mass distribution of the BH population was not affected much by most initial parameters, with the  exception of metallicity. As expected, the highest-mass BHs were found in models with the lowest metallicity. In general, BH masses were found to be within the range $7\, M_{\odot}$ to  $25\, M_{\odot}$
for both single and binary BHs. Moreover, single BHs set the overall shape of the BH mass distribution.

B04 also found that the orbital periods for the binaries in their models cover a broad range, $P_{\rm orb}\sim0.1 - 10^6\,$d. The distribution had two distinct peaks, one at small periods, centered around $10\,\rm days$, and another around $10^6\,\rm days$.
 Moreover the orbital period distribution, like the mass distribution, was dependent on metallicity. With increasing metallicity the shorter-period BH binaries were suppressed, while more of the longer-period
 binaries survived.

 \citet[][hereafter B06]{Bel06} extended the population synthesis approach of B04, by including the possibility that BHs could be ejected from their original star cluster, depending on the recoil speed following  SN explosions.
They varied the assumed cluster potential and escape speed in their models over the full relevant range, taking into account that  smaller clusters have escape speeds as low as $\sim 10\,\rm km\,s^{-1}$, while
the largest clusters have escape speeds as high as $\sim 100\,\rm km\,s^{-1}$.
B06  found that a significant fraction of BHs could be ejected from their cluster. At early times the number of BHs ejected increased with time as progressively less massive stars formed BHs which received larger kicks, removing them from the cluster.
 At later times ($>15\,\rm Myr$), no more BHs were ejected.
As previously seen in B04, the orbital period of the retained BH population was bimodal, with only short-period binaries being ejected from their cluster (long-period binaries are more prone to disruption).

\subsection{Ultraluminous X-ray Sources (ULXs)}

Observations of extragalactic ULXs can be interpreted in a number of ways. In this section we focus on the possible implications of the various interpretations for BHs in clusters.

The extremely high luminosities of some sources exceed the Eddington luminosity of most stellar mass BHs and hence are incompatible with simple models of BH XRBs assuming isotropic emission. Anisotropic or beamed emission has been discussed extensively in the literature as a possible explanation for this apparent super-Eddington luminosity. \citet{Kin01} have argued that ULXs might correspond to a phase of rapid mass transfer during the lifetime of the XRB. They have also demonstrated that this assumption along with mild beaming can explain ULXs with luminosity $L_{\rm X}<10^{40}\,\rm ergs\, s^{-1}$. Still there are some difficulties in explaining ULXs with luminosity $L_{\rm X}>10^{40}\,\rm ergs\,s^{-1}$ (as a BH of moderate mass requires extreme beaming fractions to explain this high a luminosity). In addition, this accreting BH scenario requires a very massive companion ($q=M_2/M_1>1$, where $M_1$ is the mass of the BH and $M_2$ is the mass of the companion).
Finally, there have been  observations which claimed that emission from the ULXs might be isotropic (an isotropic nebula was observed around the ULX in Holmberg II \citep{Pak01,Kaa04} and the ULX $\rm M81$ X-9 \citep{Mil04}). Thus, whether or not anisotropic emission can explain all ULXs remains quite uncertain.

Another suggestion from the X-ray spectrum is that the brightest ULXs observed in young starburst environments might harbor a more massive BH with $M_{BH}\sim10^2-10^3\, M_{\odot}$ \citep{Col99,Mus00,Van03, Mil04}, a so-called intermediate-mass BH (IMBH). Theoretically, runaway collisions during an early episode of core collapse  \citep{Gur04,Por04,Fre06,San12}
provide a way to form IMBHs in young star clusters. An IMBH formed in the dense cluster environment can dynamically acquire a binary companion, but subsequent mass transfer leading to ULX formation is expected to
occur only rarely \citep{Ble06}.
The steep hard X-ray emission of the ULXs observed, is found to be consistent with that of an accreting black-hole in its soft (high) state \citep{Fab06}. Observations of the accretion disk spectra also provide evidence for the existence of IMBHs \citep{Mit84,Mil03,Mil04}.

However, high mass ($\sim80\,\rm M_{\odot}$) stellar BHs formed in low metallicity environments also explain observations of some ULXs \citep{Gon06,Ro06,Cop07,Map09,Zam09}.
\citet{Map11} investigated the formation of massive stellar BHs in star clusters. According to stellar evolution calculations \citep{Fry99,Heg02,Heg03}, a star with a final mass $M_{\rm fin}>40\, M_{\odot}$ immediately before collapse could avoid a SN explosion and directly collapse to a BH of almost the same mass.
 \citet{Map11} simulated star clusters with a Salpeter IMF and King density profiles. They introduced a massive stellar BH (MSBH) with a binary companion in their simulations assuming that such a massive BH had formed in the cluster at sufficiently low metallicity. They assumed that the MSBH should be in a binary since both its progenitor and the MSBH itself were among the most massive stars in the cluster \citep{Kul93}. They chose the mass of the MSBH to be a constant $50\, M_{\odot}$. Moreover, the mass of the binary companion was chosen to be in the range $10<M_{\rm companion}<50\, M_{\odot}$ in order to explain ULX luminosities \citep{Pat05} with stable mass transfer \citep{Rap05}. The separation a ($0.1<a<10\,\rm AU$) 
 was chosen such that the binary was not easily ionized during the evolution of the cluster.
In stead they found that a large number of MSBHs were ejected from the cluster in less than $10\,\rm Myr$ because of close interactions. Moreover, the MSBHs ejected from the cluster retained their binary companions. Previous population synthesis studies like \citet{Lin10} suggested that MSBHs could hardly become Roche lobe overflow high mass XRBs due to the absence of natal kicks in their formation pathway. However, dynamical studies in \citet{Map11} suggested that these MSBHs formed through direct collapse could pass through Roche lobe overflow high mass X-ray binary phase since dynamical interactions had similar effects as natal kicks. \citet{Map11} concluded that dynamical interactions changed the orbital parameters of these MSBHs, favoring the occurrence of mass transfer.

  \citet{Map13} investigated the evolution of young clusters ($\sim100\, \rm Myr$) using direct $N$-body simulations. They explored $3000-4000\, M_{\odot}$ clusters with varying metallicities. Their models included metallicity-dependent stellar evolution recipes \citep{Hur00}, binary evolution using the {\tt Starlab} code \citep{Por96}, and natal kicks imparted to the stellar remnants as implemented in {\tt Starlab} \citep{Por96}. They found that three-body encounters, and especially exchange interactions, play an important role in the evolution of the massive ($>25\,M_{\odot}$) BHs in all environments. Almost $75\,\%$ of the massive ($>25\,M_{\odot}$) BH population that were in a binary at some point, acquired a binary companion through dynamical exchanges. However, for lower mass BHs, this percentage was reduced to $20\%$. Moreover, they also found that all the BHs with a companion overflowing its Roche lobe, acquired their companions through dynamical exchange interactions.
Clearly, these results suggest that a complex interplay between binary evolution and cluster dynamics is responsible for the formation of XRBs in these systems.

\section{Numerical Simulations}

To study the dynamical evolution of BHs in young star clusters we have used our Cluster Monte Carlo (CMC) code, based on the classic work of \citet{Henon}, as implemented and described in \citet[][and references therein]{Fre07}. To model stellar evolution CMC employs the single star evolution formulae of  \citet{Hur00} and the binary star evolution formulae of  \citet[BSE]{Hur02}. We have made some modifications to the formulae used for determining the masses of compact remnants and to stellar wind prescriptions, which we outline in Sections~3.3.1 and~3.3.3. We also note that \citet{Hur02} derived their fitting formulae based on detailed models for stellar masses up to $100\, M_{\odot}$. We have extended this to include stars up to $150\, M_{\odot}$ in some of our simulations, covering the full range of theoretically predicted stellar masses from star formation (Weidner \& Kroupa 2004).

\subsection{Initial Conditions}
Our basic cluster model starts with a King model with $W_0 = 5$ and initial $N$ in the range $2\times10^5-5\times10^5$ (typical to the extragalactic young massive clusters). In our simulations, we vary the initial binary fraction from $f_{b}=0\%$ to $f_{b}=50\%$.
A binary system is described initially by four parameters: the mass of the primary ($M_{1}$) (the initially more massive component), the mass ratio $q= M_2/M_1$, where $ M_2$ is the mass of the secondary (initially less massive component), the semi-major axis $a$ of the orbit, and the orbital eccentricity $e$. We assume that the initial distributions of these parameters are independent. 
For both single stars and binary system primaries we adopt the Kroupa initial mass function (IMF) in the mass range $M_1= 0.08-100\,  M_{\odot}$. The masses of the secondary stars in the binary systems, are sampled from the mass ratio $q$ assumed to be constant between $0$ and $1$.
We have chosen two different initial binary separation distributions for this study, namely, (i) \citet[similar to that assumed in the studies of B04 and B06]{Iva05} which includes many soft binaries and (ii) \citet{Hur05} prescription.
For eccentricities we assume a thermal distribution initially.

In our models we have also considered a range of metallicities: $Z=0.0002$, $Z=0.001$, and $Z=0.02$ (solar).  
For our first set of simulations (Table \ref{t1}) we have used a metallicity $Z=0.001$, typical of old globular clusters. 
We consider two ``standard'' models in this set: model~A and model~B, both initiated a with $N=5\times10^5$ and binary fraction $f_{b}=50\%$.
The number of BHs formed in a cluster is a function of the initial number of stars in the cluster ($N$) and the IMF, and in our cluster simulations (with no primordial binaries and typical Kroupa IMF) we expect $N_{\rm BH}=5\times10^{-4}N$ BHs to form after $10\,\rm Myr$.
For these set of simulations we have used the orbital period distribution \citep{Iva05}, wind mass loss prescription and the prescription for the calculation of masses of remnant objects (\citet{Bel08}, discussed in more detail in Sec.~3.3.3), similar to B06 for easier comparison.
Model A represents a dense cluster ($\rho_c(0)=1.3\times10^5\,M_\odot\,{\rm pc}^{-3}$)
with a virial radius ($r_v$) of $1.25\,\rm pc$. 
Model~B on the other hand represents a zero-density cluster, and we have studied model~B with a pure population synthesis approach, i.e., explicitly turning off dynamics in our simulations. In Tables~\ref{t4} and~\ref{t5} we have listed the subpopulations of all BHs (both single BHs and BHs in binaries) for models~B and~A, respectively. For rest of our model simulations in Table \ref{t1}, we vary the initial central density from as high as $\rho_c(0)=1.3\times10^5\,M_\odot\,{\rm pc}^{-3}$ to $\rho_c(0)=2\times10^3\,M_\odot\,{\rm pc}^{-3}$ for different initial $N$.

A strong correlation between metallicity and the number density low mass X-ray binaries, LMXBs ($L_{X}>10^{36}\,\rm  erg\,s^{-1}$),  in GCs, has been discussed for Milky Way, M31 \citep{Bel95} and NGC 4472 \citep{Kun02}. It has also been observed that the metal poor Large Magellanic
Cloud (LMC) has a lower ratio of LMXBs to high mass X-ray
binaries (HMXBs) than the Milky Way which is metal rich \citep{Cow94,Ibe97}.
Furthermore, ULXs have been observed in environments with a wide range of metallicities. 
Winter et al.\ (2007) found solar abundances for the ULXs (in 11 nearby galaxies, spiral galaxy NGC
4559 and in
M33) they investigated with XMM-Newton spectra. 
However, a fraction of ULXs are often associated with extremely low metallicity environments, especially in galaxies with high specific star formation rates \citep{Pak02,Zam04,Sor05,Swa08,Map09,Map10,Pre13}. Hence, in our second set of simulations (Table \ref{t2}), we compare the primordial BH populations 
for solar metallicity to those in extremely low metallicity clusters (with $Z=0.0002$).
For the simulations in Table \ref{t2} with low metallicity, as well as with solar metallicty, we have adopted the \citet{Bel10} treatment of compact object formation. All the models have been initiated with $N=5\times10^5$ stars varying the initial core density ($\rho_c(0)\sim1.3\times10^5-5\times10^3\,M_\odot\,{\rm pc}^{-3}$) and the primordial binary fraction ($f_b\sim0-50\%$).

Finally, we considered cluster models with a higher binary fraction for massive stars, as in the Wd~1 cluster (Table \ref{t3}). Wd~1 is the most 
massive young clusters in our Galaxy. The binary fraction in this cluster is thought to be 100\% among massive stars, with masses 
above $45\, M_{\odot}$ \citep{Cla08}. In our models we assume that all stars with mass $>45\, M_{\odot}$ are initially in binaries. 
The binary fractions quoted in Table \ref{t3} for these simulations refer to the binary fraction of less massive stars ($<45\, M_{\odot}$). Here our goal was to study the effects of the higher binary fraction for massive stars. We used a metallicity of $Z=0.001$ and  the \citet{Hur05} orbital period distribution. All other initial conditions are the same as in our first set of simulations. 
Here also we consider two representative models, W1 with high density ($\rho_c(0) \sim1.3\times 10^5\,M_\odot\,{\rm pc}^{-3}$) and W2, with no dynamics.

\subsection{Comparison with Population Synthesis}

In this paper we study the initial population of BHs in dense star clusters with full dynamics, along with a realistic stellar evolution model. 
Previous studies like those of B04 and B06 have been done with a population synthesis approach, without taking into account the important role played by
 stellar dynamics in determining the numbers and characteristics of BHs formed in dense clusters.

B04 and B06 used a Kroupa IMF with mass spectrum from $4\, M_{\odot}$ to $150\, M_{\odot}$. A higher minimum mass in the IMF simply saves computational time 
as the low-mass end of the IMF does not affect the population synthesis results. However, in dynamical simulations the cluster potential is computed by summing 
up the potential due to each star. Since it is the lower mass stars that dominate the total mass, these stars will also dominate the cluster potential and very much affect the dynamics. For that reason, in our simulations we take into account the low-mass end of the IMF and adopt a standard Kroupa IMF with $M_{\rm min}=0.08\, M_{\odot}$ and $M_{\rm max}=100\,M_{\odot}$, representative of real clusters.
 
B06
did not have to explicitly consider models with different binary fractions since their results could be easily generalized by simply weighing differently the numbers obtained for a population of single stars and for a population of binaries. In contrast, in our dynamical simulations for dense clusters, binary interactions play a major role and results cannot be rescaled: the primordial binary fraction in our simulations must be set as an initial condition for each model. We will show results for specific primordial binary fractions $f_b$ ranging form $0\%$ to $50\%$.

  \subsection{Stellar Evolution Assumptions Affecting Compact Objects}
  
   Our Monte Carlo code includes all necessary physics such as two-body relaxation, an explicit treatment of all stellar collisions, and direct integration of close encounters using Fewbody \citep{Fre04}. In this section we discuss the modifications that we have implemented in BSE.
  
 \subsubsection{Masses and Radii of Remnants}
 
 An important feature of BH evolution in clusters that impacts heavily on their stellar dynamics is the remnant mass function.
The remnant masses of NSs and BHs in this paper are calculated as in \citet{Bel08}, whose method we briefly summarize here.

\citet{Bel08} determine the masses of NSs and BHs by using information on the final CO ($M_{\rm CO}$) and FeNi  ($M_{\rm FeNi}$) core masses, 
from the results of detailed simulations by \citet{Hur00} and \citet{Tim96}, combined with the knowledge of the pre-SN mass of the star. 
For a given initial ZAMS mass ($M_{\rm ZAMS}$), $M_{\rm FeNi}$ is obtained as follows

\[
M_{\rm FeNi} = \left\{ 
  \begin{array}{l l}
    $1.50$ &  \text{$M_{\rm CO} < 4.82$ ($M_{\rm ZAMS} < 18\, M_{\odot}$),}\\
    $2.11$ &  \text{$4.82\le M_{\rm CO} < 6.31$ ($18\, M_{\odot} \le M_{\rm ZAMS} < 25\, M_{\odot}$}),\\
    \text{$6.9 M_{\rm CO}-2.26$}  &  \text{$6.31\le M_{\rm CO} < 6.75$ ($25\, M_{\odot} \le M_{\rm ZAMS} < 30\, M_{\odot}$),}\\
    \text{$0.37 M_{\rm CO}-0.07$}  &  \text{$M_{\rm CO} \ge 6.75$ ($M_{\rm ZAMS} \ge 30\, M_{\odot}$).}\\

  \end{array} \right.
 \]

For sufficiently large progenitor masses, some of the envelope material may not be ejected in the SN explosion and instead falls
 back onto the newly formed remnant. 
The mass fraction of the stellar envelope falling back is denoted by $\xi_{fb}$ in this paper.
 For stars with $M_{\rm ZAMS} \le 20\, M_{\odot}$, $\xi_{fb}=0$ and only the stellar core influences the resultant remnant mass. For stars with $M_{\rm ZAMS} \ge 42\, M_{\odot}$, the whole pre-SN star is assumed to collapse directly and form a BH, i.e., $\xi_{fb}=1$. Stars in the intermediate mass range 
 with $20\,\rm M_{\odot}<M_{\rm ZAMS} < 42\, M_{\odot}$
undergo core collapse with $0<\xi_{fb}<1$. The point to note here is that the outcome of the core collapse depends on the collapsing core and not on $M_{\rm ZAMS}$. 

For standard wind mass loss assumptions \citep{Hur02} with solar metallicity, the masses of the compact objects are obtained as,
\[
  M_{\rm rem,bar} = \left\{ 
  \begin{array}{l l}
    \text{$M_{\rm FeNi}$} &  \text{$M_{\rm CO} < 5\, M_{\odot}$,}\\
    \text{$M_{\rm FeNi}+\xi_{fb}(M_{\rm fin}-M_{\rm FeNi})$} &  \text{$5\,  M_{\odot}\le M_{\rm CO} < 7.6\,  M_{\odot}$},\\
   \text{$M_{\rm fin}$}  &  \text{$M_{\rm CO} \ge7.6\, M_{\odot}$}.\\
   
  \end{array} \right.
\]
The mass ranges for ``no fall back'' ($\xi_{fb}=0$), ``partial fall back''
 ($0<\xi_{fb}<1$), and direct collapse ($\xi_{fb}=1$) are estimated from core collapse SN simulations by Fryer (1999) and the subsequent analysis of Fryer \& Kalogera (2001).

The gravitational mass of a NS is obtained from the baryonic mass using the expression derived by  \citet{Lat89}, 
\begin{equation}
M_{\rm rem,bar}-M_{\rm rem}=0.075 M_{\rm rem}^2
\end{equation} 
and the radius of all NSs is simply set to $10\,\rm km$. The gravitational mass of a BH is set to  
$ M_{\rm rem}=0.9\, M_{\rm rem,bar} $
and the BH ``radius'' is assumed to be the Schwarzschild radius, $R_{\rm BH}=2GM_{\rm BH}/c^2$.

As implemented in \citet[see their section~2.3.1]{Bel08} we also allow low-mass NSs to form via the capture of electrons onto $^{24}{\rm Mg}$, $^{24}$Na, and $^{20}$Ne nuclei within ``electron capture supernovae'' (ECS), which affects the low-mass end of the remnant mass function.  Moreover the masses of compact object can increase through accretion in binary systems as discussed in \citet[][see their section~5.7]{Bel08}.
In our standard model, to facilitate comparison with \citet{Bel08}, we define a BH to be any compact object with mass exceeding the maximum NS mass, $M_{\rm max,NS}=3\, \rm M_{\odot}$. For completeness we also use $M_{\rm max,NS}=2.5\,  M_{\odot}$ in some simulations.

\subsubsection{Supernova Kicks}

During SN explosions 
the remnant may receive a significant velocity kick due to asymmetries in the explosion \citep{Lyn94}. This can be of considerable importance to a star in a binary, as the kick imparted to the remnant might result in disrupting the binary. It has been argued that all NSs, and possibly also BHs (see, e.g., \citet{Gua05,Fra09,Jan13}) 
receive such natal 
kicks. In our simulations we give natal kicks to both NSs and BHs during SNe, following the prescriptions in \citet{Hur02} and \citet{Kie09}. For better comparison with B06 
we have chosen to use the \citet{Arz02} kick velocity distribution in our standard models~A and~B. This is an empirical distribution based on the observed 
proper motions of single radio pulsars in the field. 
It is a two-component velocity distribution with characteristic velocities of $90$ and $500~{\rm km~s}^{-1}$. For rest of the simulations in this paper we have used 
the \citet{Hob05} distribution, which is a Maxwellian with standard deviation $\sigma = 265~{\rm km~s}^{-1}$. 
Both NS and BH kicks are drawn from these distributions; however, only compact objects formed with ``no fall back'' receive full kicks, while for compact objects receiving ``partial fall back''  we limit any kick velocity ($v_k$) drawn from these distributions by the mass fraction of the stellar envelope that falls back ($v_k=(1-\xi_{fb})\times v_k$).

\subsubsection{Wind Mass Loss}

For our standard models and for simulations with $Z=0.001$ we have used the wind mass loss prescriptions from \citet{Iva05} and \citet{Hur02}, respectively. For our  simulations with $Z=0.0002$ we have modified the wind prescription as done in \citet{Bel10}, based on the work of \citet{Vin01}. With this modified wind prescription in their simulated cluster models (using population synthesis), \citet{Bel10} found that very massive BHs (with masses up to $\sim 100\, M_{\odot}$) 
could be formed at these extremely low metallicities.

\section{Results}
\subsection{Properties of Single and Binary BHs}
B06 investigated the evolution of single and binary BHs formed in a cluster with a population synthesis approach (but taking into account ejections of BHs from their parent cluster by SN kicks), using the StarTrack code. Here we first examine whether we can reproduce some general results of B06 using BSE with our updates to its stellar evolution prescriptions. 

\begin{figure}[ht]
\begin{center}
\includegraphics[clip, width=10.6 cm ]{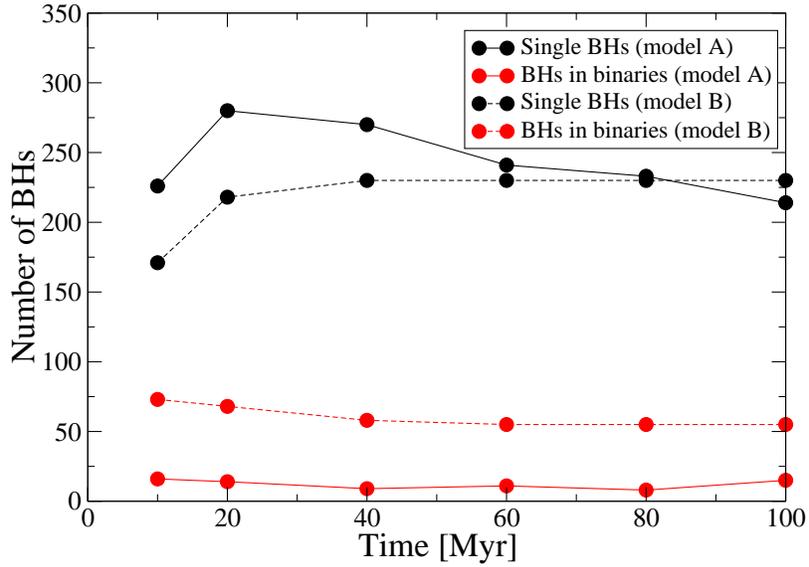}
\end{center}

\caption{\label{fig:allsinbin}Total number of single BHs (black symbols) and binary BHs (red symbols) formed in the cluster, for our standard models A and B. Model A (solid lines) represents a dense cluster (virial radius, $r_v=1.25\,\rm pc$) and model B (dashed lines) represents a zero-density cluster (simulated with a population synthesis approach).}

\end{figure}

Figure \ref{fig:allsinbin} shows the total (both retained and ejected) number of single and binary BHs formed in a cluster with $N=5\times10^5$ stars and a $50\%$ primordial binary fraction, as the cluster evolves until the end of the simulation at $100\,\rm Myr$. Model~B represents a low-density cluster in which dynamics does not play a significant role. We have investigated model~B with an approach similar to B06, for easier comparison. 

Clear agreement between B06 and our model~B is seen for the trend in the evolution of the total number of BHs over time. Figure \ref{fig:allsinbin} shows that the total number of single BHs increases over time during the initial $20\,\rm Myr$, while the number of binary BHs decrease. The binary BHs decrease in number because of stellar evolution events such as SN explosions, which tend to disrupt binaries, and CEs, 
which can make binaries merge. Figure \ref{fig:allsinbin} also shows that, as the cluster evolves beyond $20\,\rm Myr$, the total number of single BHs and BH binaries becomes constant (as in B06) because the cluster is depleted of the massive progenitor stars which form BHs. 

Next, we want to compare these population synthesis results with our dynamical simulations for a dense cluster (see Table \ref{t4} and Table \ref{t5}).
Model~A represents a dense star cluster in which stellar dynamics is important (in contrast to model~B). 
One of the results of including stellar dynamics with regards to BH numbers, is that at late times ($>20\,\rm Myr$) the total (retained and ejected) number of single and binary BHs is no longer constant (see Figure \ref{fig:allsinbin}, solid lines). Moreover, we also observe an increase in the population of single BHs (by $\sim10\%$ compared to model~B) and a decrease in the number of BHs in binaries, since all wide binaries are disrupted to form single BHs in dense cluster environments.

\begin{figure}[ht]
\begin{center}
\includegraphics[clip, width=10.6 cm ]{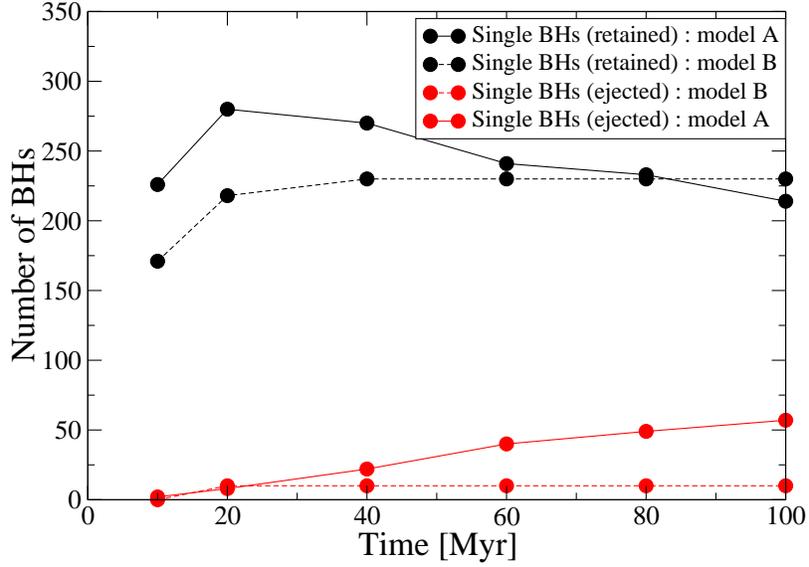}
\end{center}

\caption{\label{fig:allsin}Number of single BHs retained in the cluster (black symbols) and ejected from the cluster (red symbols) within different time intervals for our standard models. Model~A (solid lines) represents a dense cluster and model B (dashed lines) represents a zero-density cluster (simulated with a population synthesis approach).}
\end{figure}

\begin{figure}[ht]
\begin{center}
\includegraphics[clip, width=10.6 cm ]{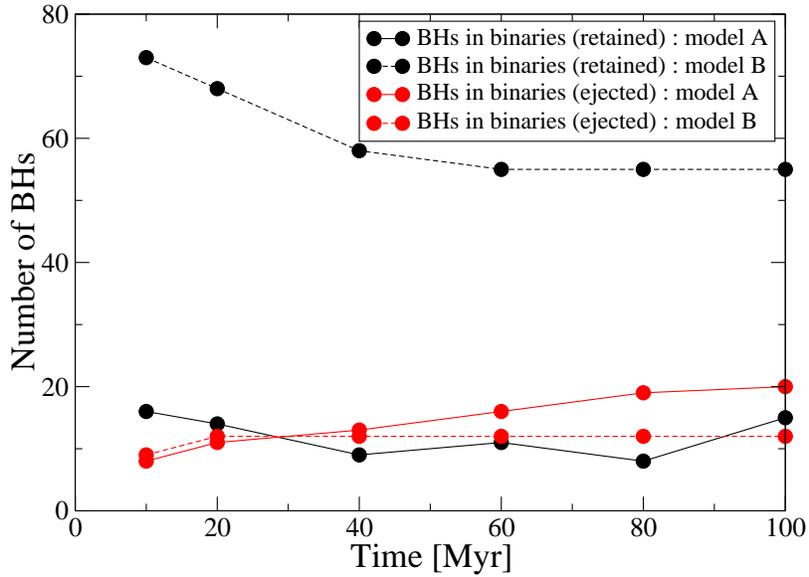}
\end{center}

\caption{\label{fig:allbin}
Same as Fig. \ref{fig:allsin} but for BHs in binary systems. }

\end{figure}

Also from Figure \ref{fig:allsinbin}, we see that the early phase (upto $20\,\rm Myr$) of evolution of  a dense cluster mimics a low-density cluster, with respect to the increasing trend in the number of BHs.  This shows the dominant importance of stellar and binary evolution at early times.  After $20\,\rm Myr$, the number of single BHs and BHs in binaries in model~A never becomes constant. Clearly dynamical interactions are responsible for these differences in the later phases of evolution. To investigate this further we now examine separately the retained and ejected populations of single BHs (Figure \ref{fig:allsin}) and binary BHs (Figure \ref{fig:allbin}), in more detail, for our two standard models.

Figure~\ref{fig:allsin} shows the number of single BHs retained in (black symbols) and ejected from (red symbols) the cluster for models~A and~B. Again we 
compare the results from model~B with those in B06, 
and then contrast them with model~A, where dynamics is important. 
In model~B, a small fraction ($\sim 4\%$) of single BHs are ejected from the cluster, and only within the first $20\,\rm Myr$. The ejected single BH population results from the single BHs in the cluster receiving SN kicks, and from the disruption of binary BHs through SNe similar to B06. In contrast, for model~A, single BHs are ejected during the entire evolution of the cluster (up to $\sim100\,\rm Myr$). Initially, single BHs are ejected mainly through SN kicks but at later times, as the BHs mass segregate, more single BHs are ejected ($\sim 80\%$ of the ejected population) through recoil produced by dynamical interactions. 

Figure \ref{fig:allbin} is the same as Figure \ref{fig:allsin} but for BHs in binaries. Similar to the single BH population, BHs in binaries in model~B are ejected only through SN kicks within the first $20\,\rm Myr$, whereas in model~A, more BHs are ejected after $20\,\rm Myr$ through dynamical interactions. This explains why we see a continuous increase in the number of BHs (both single BHs and BHs in binary systems) ejected for model~A (solid red lines in Figure~ref{fig:allsin} and 
Figure~\ref{fig:allbin}) until $100\,\rm Myr$, unlike the constant number in model~B (dotted red lines). The number of single BHs and binary BH systems retained in the cluster also depends on the type of interactions occurring in the cluster. For example, dynamical interactions can increase the number of single BHs by disrupting a binary system as well as decrease the single BH population when a BH--MS system exchanges the low-mass MS star for another more massive BH. 


\subsection{Binary Ejection Fraction}

\begin{figure}[ht]
\begin{center}
\includegraphics[clip, width=10.6 cm ]{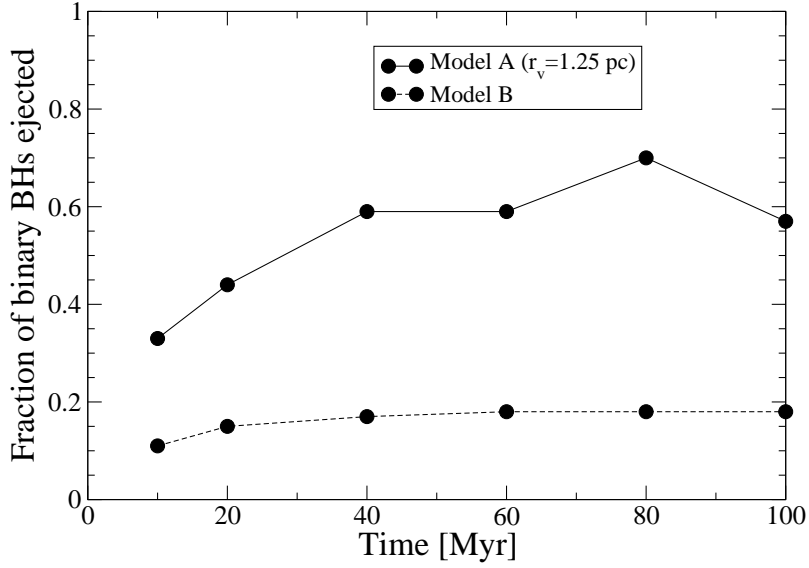}
\end{center}

\caption{\label{fig:fraccompare}Fraction of BHs in binaries ejected from the cluster within different time intervals for our standard model. The solid line represents a dense cluster (model~A) and the dotted line represents a zero-density cluster (simulated with a population synthesis approach) with negligible effects from dynamics (model~B).}

\end{figure}

We have seen in Figures~\ref{fig:allsin} and~\ref{fig:allbin} that, as a cluster evolves beyond $\sim10\,\rm Myr$, BH ejections become dominated by dynamical interactions rather than SN kicks. 
The dynamical interaction rate in a cluster depends on the central density. 
In our models we vary the initial density of the cluster by setting the initial virial radius ($r_{v}$) for the King model. 
Here we examine the dependence of the number of BHs ejected from the cluster on the density. 
In Figure \ref{fig:fraccompare}, we plot the ejected fraction of BHs that are in binaries for Model~A with $\rho_c(0)=1.3\, M_\odot\, {\rm pc}^{-3}$ ($r_v=1.25\,\rm pc$) and a the zero-density cluster.
We have included in Table \ref{t1} the BH ejection fraction from the different models, varying the initial density (from $\rho_c(0)=1.3\times 10^5\, M_\odot\, {\rm pc}^{-3}$ to $\rho_c(0)=2.0\times 10^3\, M_\odot\, {\rm pc}^{-3}$ for different N) and initial binary fraction. 
We concentrate mainly on binary BHs (binaries with one or both the stars as BHs) since these BHs will be the most massive BHs in the cluster and will sink towards the center faster than the single BHs. 
We find that the binary BH ejection fraction in a high-density cluster is significantly higher (by a factor of $\sim3$) than in a low-density cluster. 
We find that, although the BH binary ejection fraction increases with  increasing density, the primordial binary fraction does not play a significant role. 
(see also Sec.~4.6).

\subsection{Mass Distributions for Single and Binary BHs}

\begin{figure}[ht]
\begin{center}
\includegraphics[clip, width=10.6 cm ]{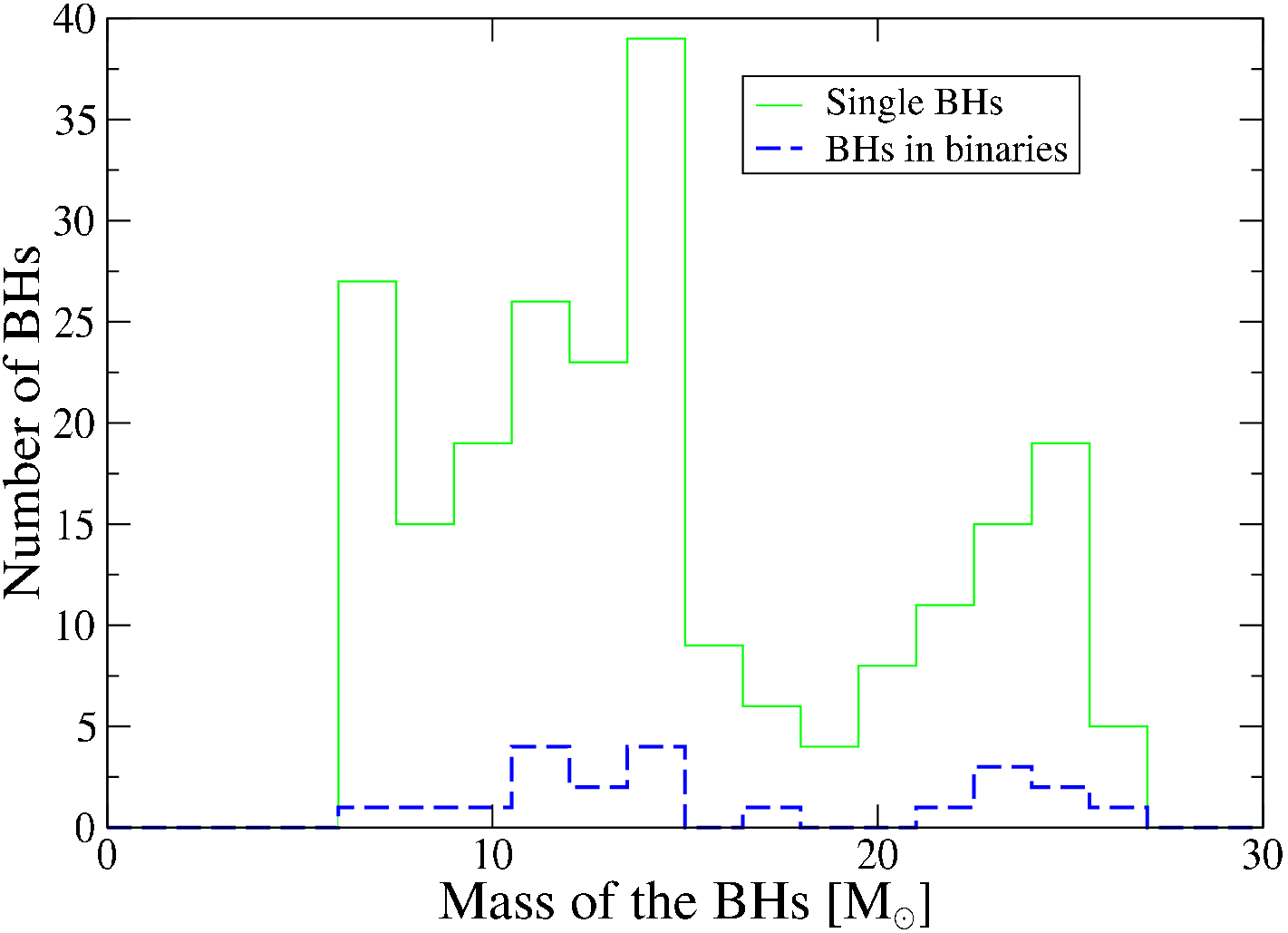}
\end{center}

\caption{\label{fig:distAall}Mass distribution of BHs in singles (green lines) and BHs in binaries (blue lines) retained in the cluster at $10\,\rm Myr$ for our standard model A.}

\end{figure}

\begin{figure}[ht]
\begin{center}
\includegraphics[clip, width=10.6 cm ]{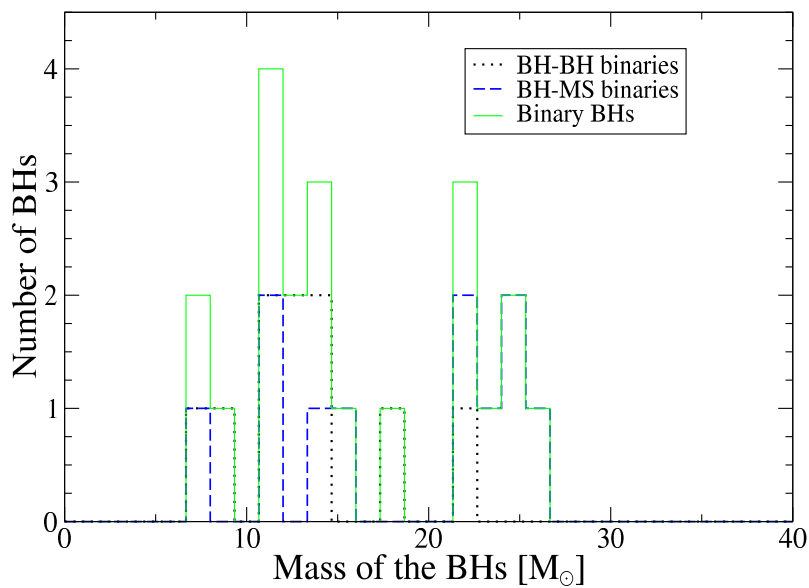}
\end{center}

\caption{\label{fig:distAbinret}Mass distribution of BHs in binaries (green solid lines) retained in the cluster at $10\,\rm Myr$ for our standard model A. Two dominant contributing system types are shown here separately:  
BH--MS binaries (red dashed lines) and BH--BH binaries (black dashed lines).}

\end{figure}

After successfully reproducing the trend in the population of BHs, as suggested in B06, with our model B, we discuss model A in more detail.
Mass distributions of single and binary BHs retained in the cluster (model A) at $10\,\rm Myr$ are shown in Figure \ref{fig:distAall}. We find that the retained single BH population has three peaks: a first peak at $6-8\,M_{\odot}$ (from stars of mass $40<M<50\, M_{\odot} $), a second peak at $10-16\, M_{\odot}$ (from stars of mass $M>50\, M_{\odot} $) and a third peak at $22-26\, M_{\odot}$ (from stars of mass $25<M<35\, M_{\odot} $). The maximum BH mass is $\sim26\, M_{\odot}$ (maximum mass is dependent upon initial metallicity, see sections 4.7 and 4.8). We also looked into the mass distribution of different binary systems (BH--BH and BH--MS) retained in the cluster at $10\,\rm Myr$ (Figure \ref{fig:distAbinret}). From Figure \ref{fig:distAbinret}, we find that at $10\,\rm Myr$ most of the binaries are in BH--MS or BH--BH systems. Moreover, BH--MS systems are in general more massive than BH--BH systems. 

\begin{figure}[ht]
\begin{center}
\includegraphics[clip, width=10.6 cm ]{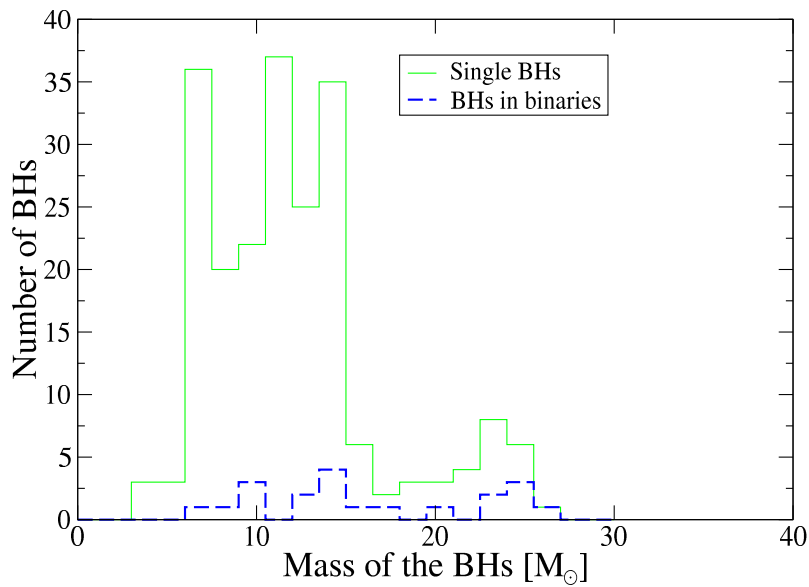}
\end{center}

\caption{\label{fig:distAbinej}Mass distribution of BHs in singles (green lines) and BHs in binaries (blue lines) retained in the cluster at $100\,\rm Myr$ for our standard model A.}

\end{figure}

In Figure \ref{fig:distAbinej}, we plot the mass distribution of BHs at $100\,\rm Myr$. Low-mass BHs formed beyond $10\,\rm Myr$, as shown in Figure \ref{fig:distAbinej}, populate the extreme low-mass end of the BH mass spectrum. We find that a fraction of BHs in the mass range $15-24\,M_{\odot}$ are missing (when compared to Figure~\ref{fig:distAall}). These BHs, being more massive, concentrate in the dense cluster core and get ejected through dynamical
interactions within $100\,\rm Myr$. 

\subsection{Orbital Period Distribution}

\begin{figure}[ht]
\vspace {0.1 in}
 \centering
 \subfigure[Model A at $10\,\rm Myr$ ]{
  \includegraphics[scale=0.30]{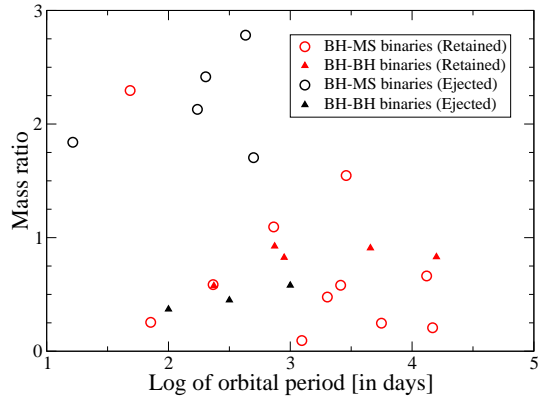}
   \label{fig:orbiA10}
   }
   
   \subfigtopskip = 0.3in	
 \subfigure[Model A at $100\,\rm Myr$]{
  \includegraphics[scale=0.30]{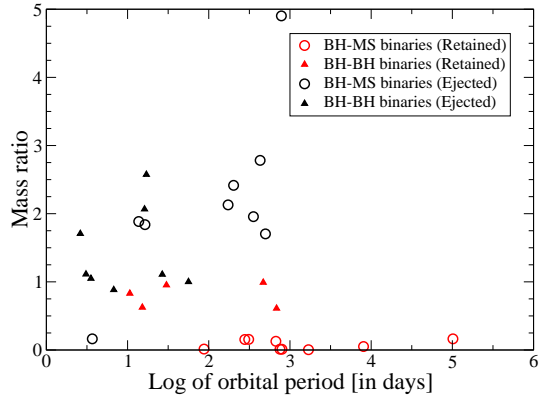}
   \label{fig:orbiA100}
   }
   
 \subfigure[Model B at $100\,\rm Myr$]{
  \includegraphics[scale=0.30]{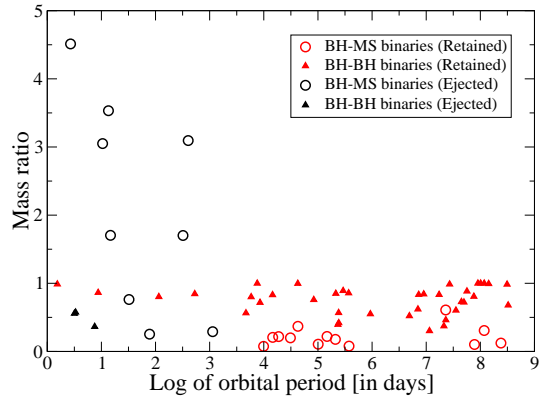}
   \label{fig:orbiB100}
   }
 \label{fig:subfigureExample}
 \caption[Optional caption for list of figures]{%
  Orbital period distribution of BH--MS systems (circles) and BH--BH systems (triangles). Systems retained in the cluster are shown in red and systems ejected are shown in black.}
\end{figure}

Figure \ref{fig:orbiA10} and Figure \ref{fig:orbiA100} show the orbital period distribution 
($P_{\rm orb}$) of BH--BH systems and BH--MS systems for model~A at $10\,\rm Myr$ and $100\,\rm Myr$, respectively. For BH--MS systems, the mass ratio shown is the mass of the MS companion divided by the BH mass. 
 For BH--BH systems, the ratio is the mass of the less massive BH divided by the mass of the more massive BH.

From Figure \ref{fig:orbiA10}, we find that the orbital period distribution covers the range $P_{\rm orb}\sim10-10^5\,\rm d$, with a peak around $P_{\rm orb}\sim10^3\,\rm d$. Systems ejected from the cluster within $10\,\rm Myr$ are mainly short-period binaries. These binaries have higher orbital speeds and hence suffer larger recoils through SN kicks during the formation of the compact object. The systems retained in the cluster at $10\,\rm Myr$ are mostly 
longer-period systems ($P_{\rm orb}>10^3\,\rm d$) with a few ($\sim 30\%$) shorter-period ones (higher mass BHs receiving low SN kicks).  

In Figure \ref{fig:orbiA100} we see that $\sim 80\%$ of the binary systems in the retained population have short periods ($P_{\rm orb}<10^3\,\rm d$). 
Moreover, all the binaries that are ejected from the cluster after $20\,\rm Myr$ through dynamical interactions are also short-period systems. Thus, in contrast to what we saw in Figure~\ref{fig:orbiA10}, there is almost no difference at $100\,\rm Myr$ in the orbital period distribution between the retained and ejected systems. The long--period ($P_{\rm orb}>10^3\,\rm d$) binaries that do exist are low-mass BH--MS binaries and reside outside the core of the cluster, while the short-period binaries 
are more massive and concentrate in the core. This demonstrates that dynamical interactions of hard binaries with single stars or other binaries tend to make them harder, a well-known result \citep{Heg75}. Figure~\ref{fig:orbiB100} shows the orbital period distribution for BH binary systems in model~B at $100\,\rm Myr$. We see that in a low-density cluster with no dynamics only $\sim 5\%$ of the BH binary systems have $P_{\rm orb}<10^3\,\rm d$ at $100\,\rm Myr$.

\subsection{ Eccentricity Distribution}

\begin{figure}[ht]
\vspace {0.1 in}
 \centering
 \subfigure[Model A at $10\,\rm Myr$ ]{
  \includegraphics[scale=0.30]{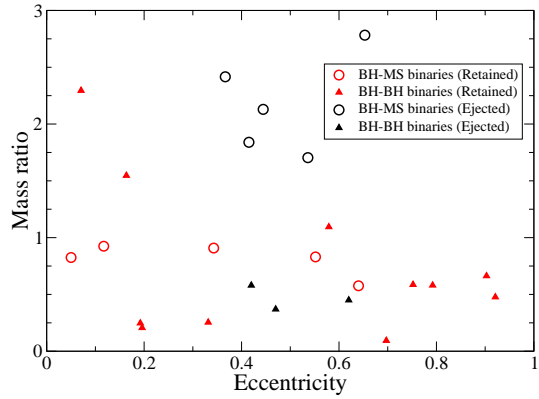}
   \label{fig:eccA10}
   }
   
      \subfigtopskip = 0.3in
 \subfigure[Model A at $100\,\rm Myr$]{
  \includegraphics[scale=0.30]{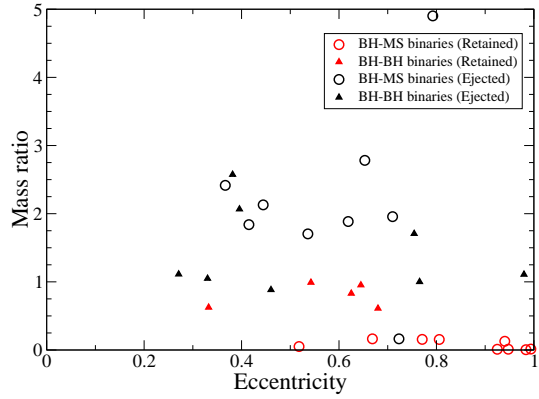}
   \label{fig:eccA100}
   }
   
         \subfigtopskip = 0.3in
 \subfigure[Model B at $100\,\rm Myr$]{
  \includegraphics[scale=0.30]{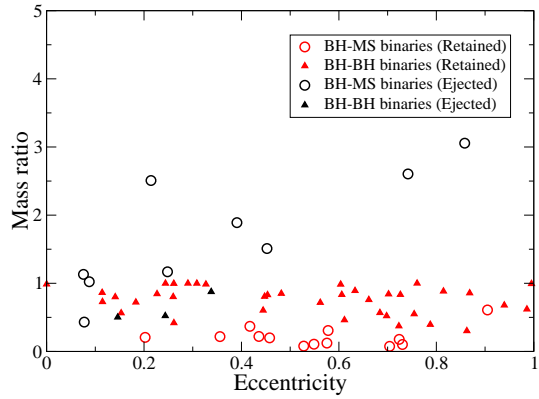}
   \label{fig:eccB100}
   }
 \label{fig:subfigureExample}
 \caption[Optional caption for list of figures]{%
  Eccentricity distribution of BH--MS systems (circles) and BH--BH systems (triangles). Systems retained in the cluster are shown in red and systems ejected are shown in black.}
\end{figure}

The eccentricity distributions for the binary systems in model A at $10\,\rm Myr$ and $100\,\rm Myr$ are shown in Figure \ref{fig:eccA10} and Figure \ref{fig:eccA100}, respectively.
At $10\,\rm Myr$ (Figure \ref{fig:eccA10}), the eccentricity ($e$) of the retained binary systems is evenly distributed between 0 and~1, while the eccentricity of the ejected population is distributed within a narrow range, $e\simeq 0.4-0.6$. However, at $100\,\rm Myr$ (Figure \ref{fig:eccA100}), we find no difference in the eccentricity distribution between the retained and ejected population (this is due to the absence of low-eccentricity systems). 
At $10\,\rm Myr$ we find $37\%$ of the retained population with $e<0.2$, whereas at $100\,\rm Myr$ there are no systems with $e<0.2$. This follows directly from the expectation that close encounters on average increase the eccentricities of binary systems \citep{Sig93,Ras95,Col03}. In Figure~\ref{fig:eccB100} we plot the eccentricity distribution for BH binary systems retained in model~B at $100\,\rm Myr$. In contrast to model~A, without dynamics, we find that $\sim 20\%$ of binary BHs have $e<0.2$.

\subsection{ Primordial Binary Fraction}

In dense star clusters binaries play a key dynamical role. A strong dynamical interaction of a binary can disrupt it, exchange one of its members for an incoming star, cause its orbit to expand or shrink, modify
its eccentricity, or cause two or more stars to physically collide. 
The dynamical friction timescale 
($t_{\rm df}$) of a star of mass $m$ is given by, 
\begin{equation}
t_{\rm df}\sim\frac{\langle m\rangle}{m}t_{\rm rh(0)}
\end{equation}
where, $\langle m\rangle$ is the average particle mass of the cluster and $t_{\rm rh}$ is the half mass relaxation time. 
A high primordial binary fraction (when initial $N$ and $r_v$  is kept constant) increases $\langle m\rangle$ of the cluster, while decreasing $t_{\rm rh(0)}$ (compare simulations  $\rm M_2$ and $\rm A$ when $N=5\times10^5$ and $r_v=1.25\,\rm pc$). However, the increase in $\langle m\rangle$ is more than the decrease in $t_{\rm rh(0)}$, increasing $t_{\rm df}$ and hence the timescale for stars to mass--segregate towards the centre of the cluster.

\begin{figure}[ht]
 \vspace{0.1 in}
 \centering
 \subfigure[Model W2 ]{
  \includegraphics[scale=0.43]{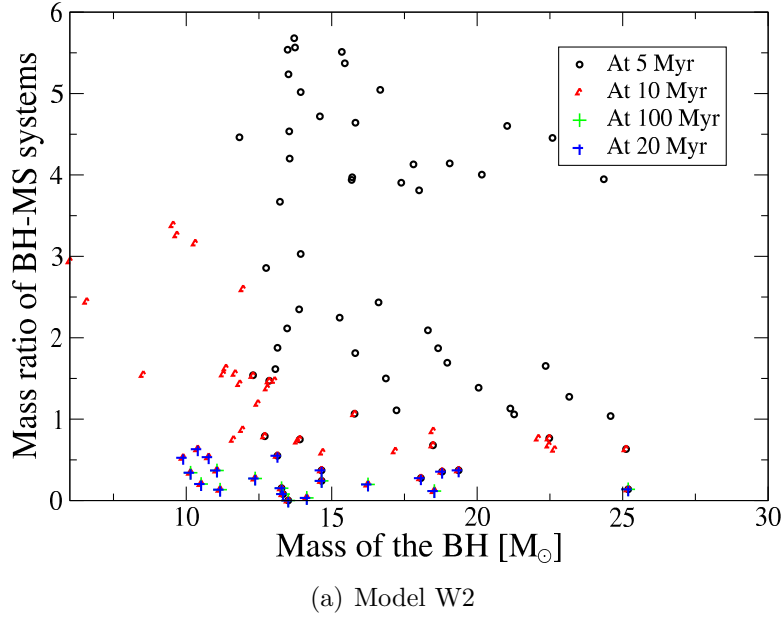}
   \label{fig:Westerlunddyn}
   }
  
           \subfigtopskip = 0.5in
 \subfigure[Model W1]{
  \includegraphics[scale=0.43]{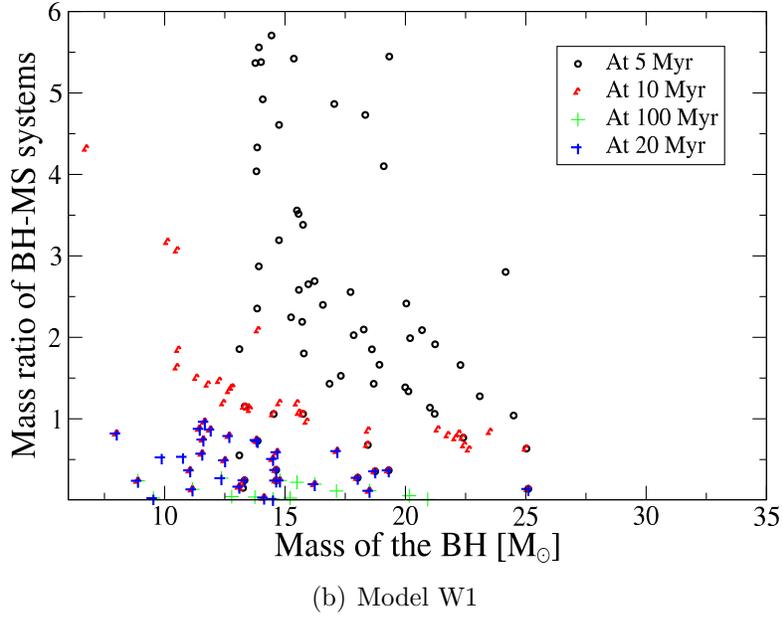}
   \label{fig:Westerlundnodyn}
   }
 \label{fig:subfigureExample}
 \caption[Optional caption for list of figures]{%
  Mass ratio of BH--MS systems in a cluster with $N=5\times10^5$ stars and $r_v=1.25\,\rm pc$ (see Table \ref{t3} for model W1 and model W2). The BH--MS systems retained in the cluster at $5\,\rm Myr$ are shown here  by black circles, at $10\,\rm Myr$ by red triangles, at $20\,\rm Myr$ by blue crosses, and at $100\,\rm Myr$ by green plus signs.}
\end{figure}

In Figure \ref{fig:Westerlunddyn} and Figure \ref{fig:Westerlundnodyn}, we focus on the effects of exchange interactions of binary systems in dense clusters. In Figure \ref{fig:Westerlunddyn}, we plot the mass ratio of BH--MS systems (at $5\,\rm Myr$, $10\,\rm Myr$, $20\,\rm Myr$ and $100\,\rm Myr$) for model W1 which is a dense cluster and in Figure \ref{fig:Westerlundnodyn}, for a zero density cluster (model W2). Comparing Figure \ref{fig:Westerlunddyn} and Figure \ref{fig:Westerlundnodyn} we find that at $5\,\rm Myr$, when cluster evolution is dominated by SN mass loss, there is no difference in the overall mass ratio of BH--MS systems in the two clusters. As the cluster becomes dominated by dynamical interactions ($>10\,\rm Myr$), the BH--MS mass ratio becomes higher in the dense cluster (Figure \ref{fig:Westerlundnodyn}). However, we find that beyond $20\,\rm Myr$, when the cluster is only left with very low mass MS stars, exchange interactions once again do not affect the ratio of BH-MS systems.

\subsection{Simulations with Metallicity $Z=0.0002$}

\begin{figure}[ht]
 \vspace{0.1 in}
 \centering
 \subfigure[At $10\,\rm Myr$]{
  \includegraphics[scale=0.43]{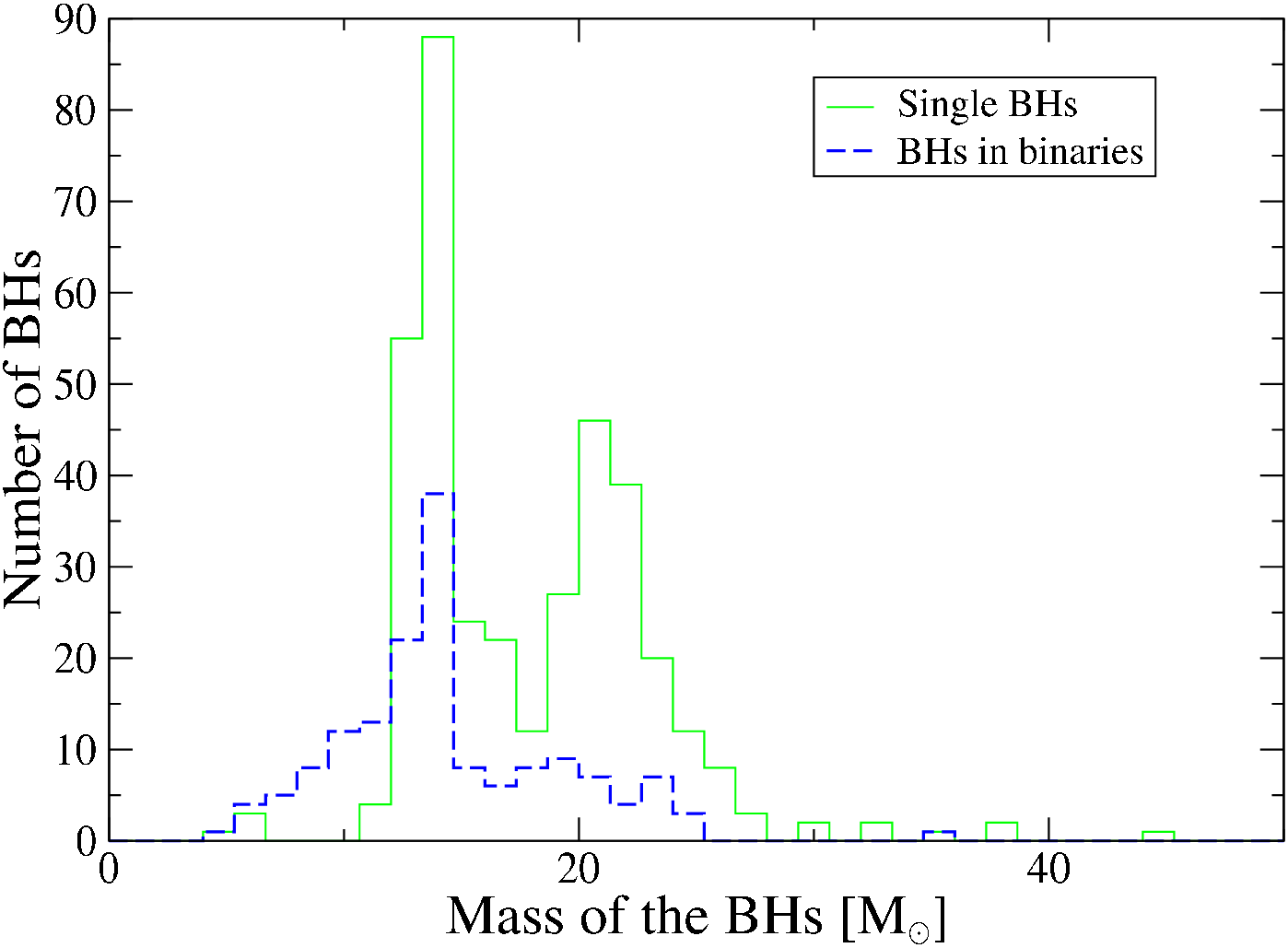}
   \label{fig:ULX10}
   }
   
              \subfigtopskip = 0.5in
 \subfigure[At $100\,\rm Myr$]{
  \includegraphics[scale=0.43]{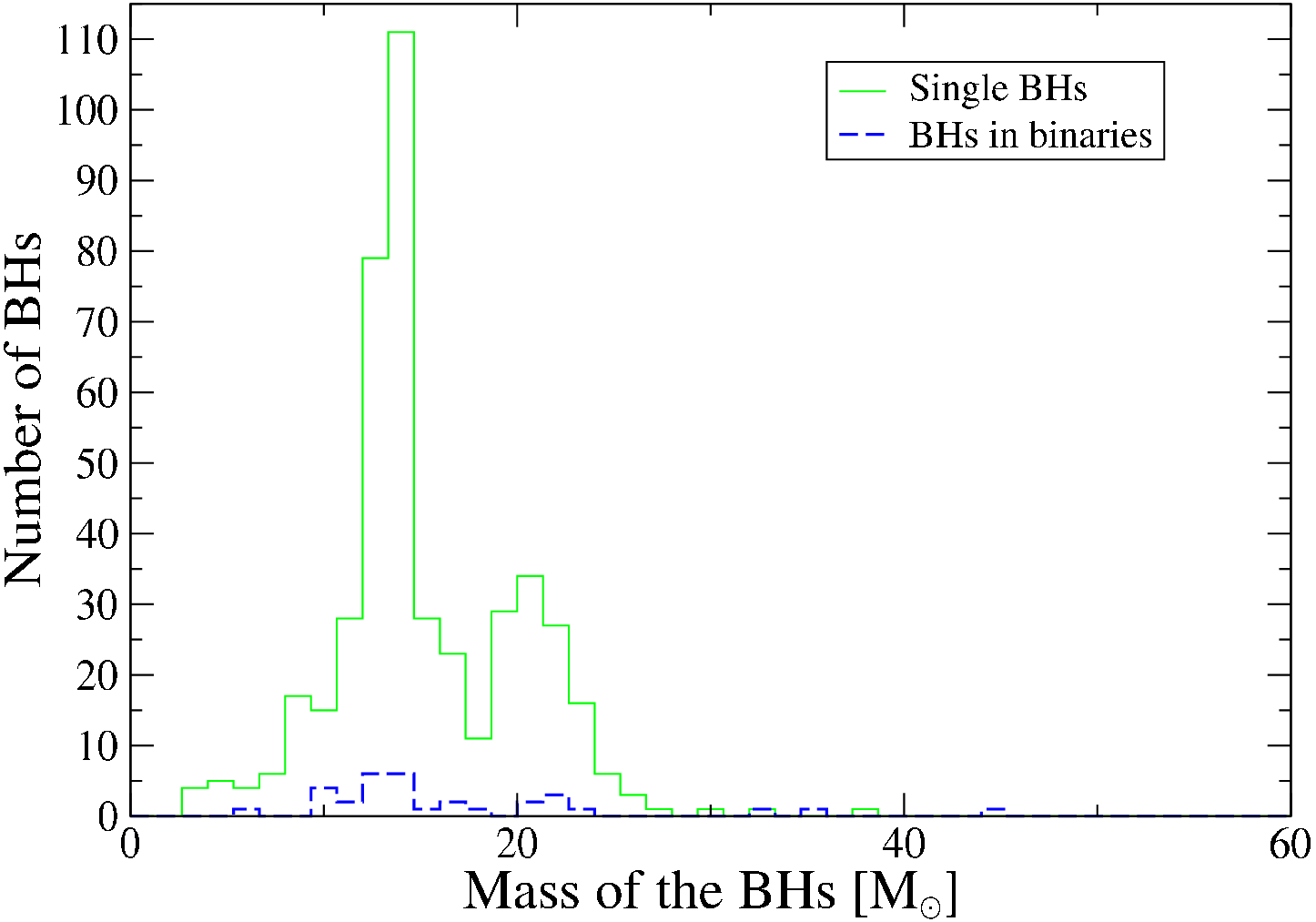}
   \label{fig:ULX100}
   }

 \label{fig:subfigureExample}
 \caption[]{%
  Mass distribution of BHs for $Z=0.0002$ with $N=5\times10^5$ stars and $r_v=1.25\,\rm pc$. Green lines represent single BHs while blue lines represent BHs in binaries.
}
\end{figure}

The overall numbers of single and binary BHs formed in the cluster at very low metallicity do not differ much from those for $Z=0.001$. However,
clear differences are seen in the mass distributions.
Figure~\ref{fig:ULX10} plots the mass distribution of single and binary BHs at $10\,\rm Myr$ for a cluster with $N=5\times10^5$ stars, $r_v=1.25\,\rm pc$ and $Z=0.0002$. We find that the maximum mass for BHs reaches $50\, M_{\odot}$ (cf.\ $M_{\rm max,BH}\simeq25\, M_{\odot}$ for $Z=0.001$). 

At $Z=0.001$, the most massive BHs are formed by stars with $M_{\rm ZAMS}\sim25-35\, M_{\odot}$, as these stars do not undergo the strong LBV winds and are only subjected to weaker metallicity dependent winds. However, at $Z=0.0002$ the progenitors of the most massive BHs are stars more massive than $50\, M_{\odot}$. In fact, the mass distribution of BHs with $Z=0.0002$ has only two peaks. Moreover, the $6-8\, M_{\odot}$ peak in the mass distribution of BHs from the high mass stars ($>50\, M_{\odot}$) found in simulations with $Z=0.001$ is absent. 

We find a dearth in the number of BHs ejected from the cluster in our simulations with $Z=0.0002$ during the initial $10\,\rm Myr$. In our simulations SN kicks are scaled with the progenitor masses of the BHs such that the more massive stars receive low magnitude SN kicks and are retained in the cluster. As the cluster evolves beyond $10\,\rm Myr$, we find that $\sim30\%$ of the BHs are ejected through dynamical interactions, among which $\sim50\%$ are in binaries. 

In Figure \ref{fig:ULX100}, we plot the mass distribution of single and BHs in binaries at $100\,\rm Myr$. We see that most of the massive ($>30\, M_{\odot}$) single BHs at $10\,\rm Myr$ have acquired a binary companion by $100\,\rm Myr$. In general, we find from our simulations that it is very common for massive single BHs to acquire a binary companion. However, the BH companion acquired in all these cases is another massive BH which has segregated to the cluster core. 

\subsection{Simulations with Solar Metallicity}

\begin{figure}[ht]
 \vspace{0.1 in}
 \centering
 \subfigure[At $10\,\rm Myr$]{
  \includegraphics[scale=0.43]{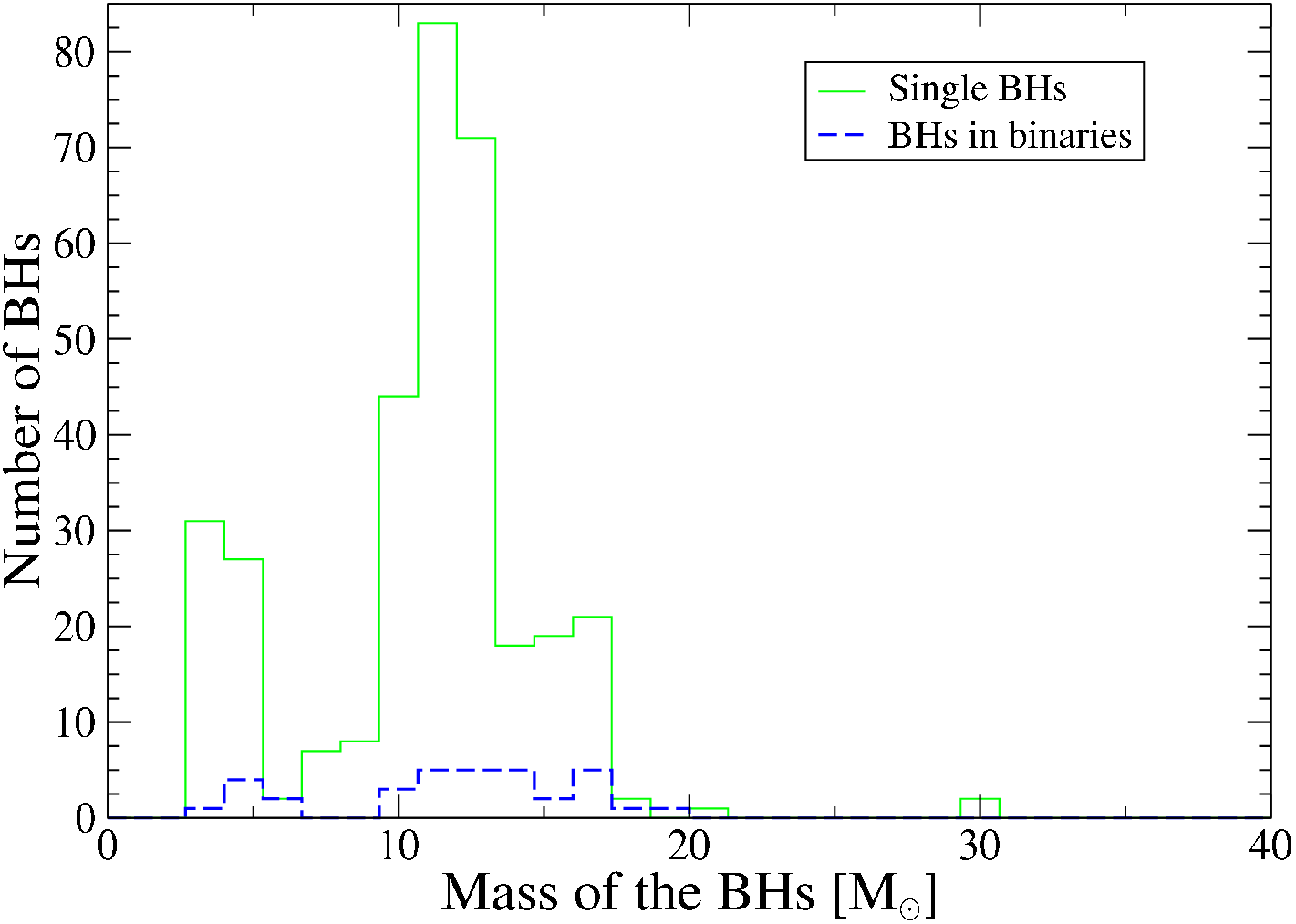}
   \label{fig:solar10}
   }
   
                 \subfigtopskip = 0.5in
 \subfigure[At $100\,\rm Myr$]{
  \includegraphics[scale=0.43]{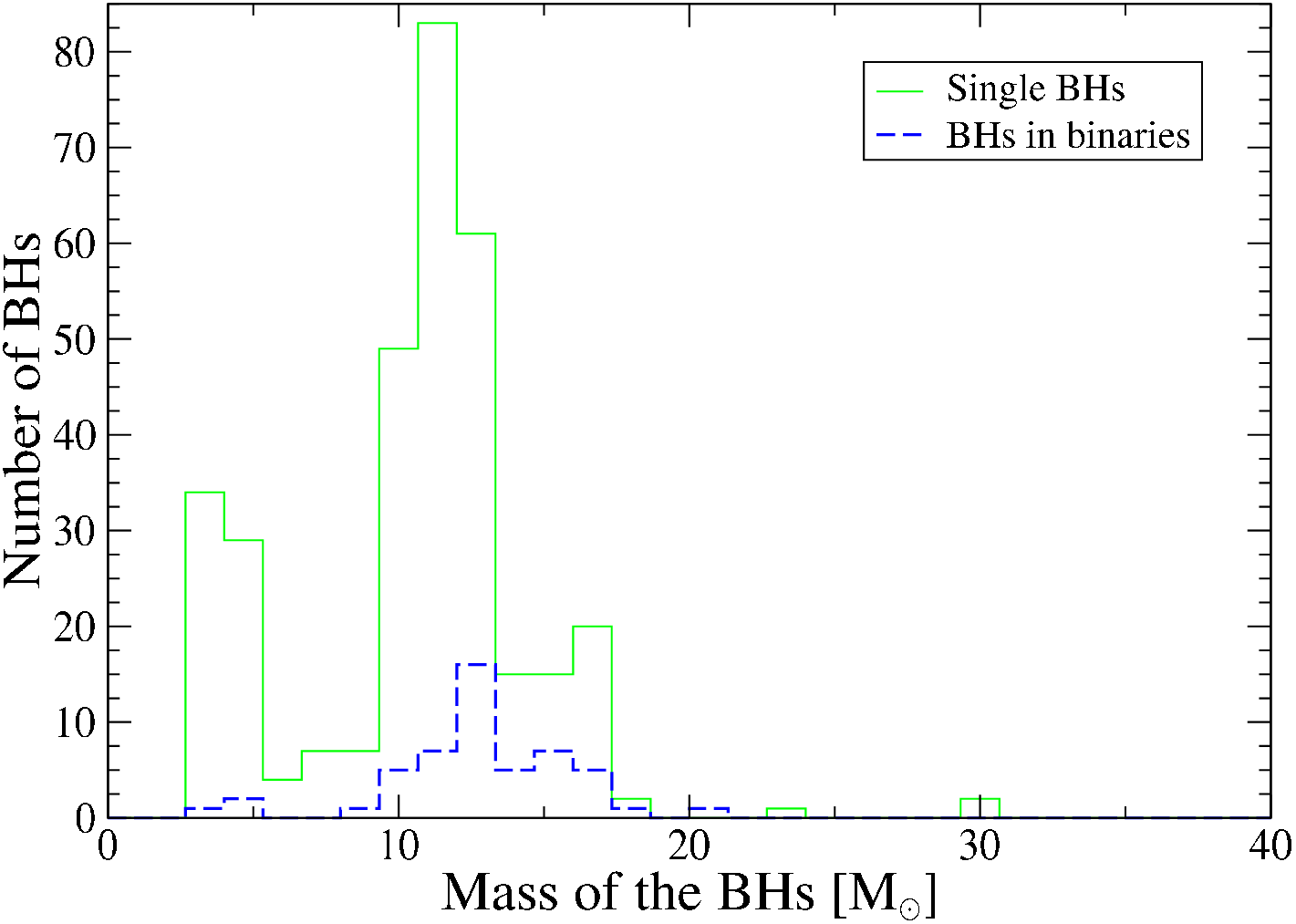}
   \label{fig:solar100}
   }

 \label{fig:subfigureExample}
 \caption[]{%
  Mass distribution of BHs for $Z=0.02$ with $N=5\times10^5$ stars and $r_v=1.25\,\rm pc$. Green lines represent single BHs while blue lines represent BHs in binaries.
}
\end{figure}

The mass distribution of BHs for our models with $Z=0.02$ at $10\,\rm Myr$ is shown in Figure~\ref{fig:solar10}.
In these simulations the masses of the BHs are systematically lower than in the population of BHs with $Z=0.001$. The BH mass distribution has two peaks and the high mass peak (around $25\,M_{\odot}$ in Fig.~\ref{fig:distAall} with $Z=0.001$) is absent.  However, in one of our simulations with $Z=0.02$, we find a single BH with a mass of $30\, M_{\odot}$; the progenitor star of this BH was formed in a collisional merger (at around $4\,\rm Myr$) during the resonant interaction of two binary systems \citep{Fre04}. 
The mass distribution of BHs at $100\,\rm Myr$ is shown in Figure~\ref{fig:solar100}. The  population of ejected BHs consists mainly of low--mass single BHs 
(with masses $\sim 10\,M_{\odot}$) and a few ($\sim 10\%$) binary BH systems (unlike what we saw in simulations for $Z=0.001$ and $Z=0.0002$). Hence, comparing Figures~\ref{fig:solar100} and~\ref{fig:solar10}, we note that some BHs in the second peak ($10-15\,M_{\odot}$ BHs) of the mass distribution are missing in 
Figure~\ref{fig:solar100}.

\section{Summary and Conclusions}

Using our Cluster Monte Carlo (CMC) 
code with full stellar evolution, we have investigated the formation and evolution of BHs in young massive stellar clusters. Our study extends and improves older work by B04 and B06 by including a full treatment of stellar dynamics. We find that, although stellar dynamics remains unimportant for the initial populations of BHs in 
low-density clusters, it can play a key role in dense clusters. Dynamical interactions between massive BH binaries and single stars or other BHs not only change the properties of these systems, but also alter the relative numbers of single and binary BHs retained in the cluster. 

During the evolution of a dense cluster (in our models, $\rho_c\sim 1.3\times10^5 \,M_\odot\,{\rm pc}^{-3}$),
 increasing numbers of BH binaries tend to get ejected from the cluster through dynamical interactions. 
In low-metallicity environments, dynamical interactions along with SN kicks, eject $\sim 60\%$ of the BH binary systems from a dense cluster within the first 
$\sim100\,\rm Myr$ of dynamical evolution. If we assume that the observable X-ray binary systems of a young massive cluster are represented by BH--MS 
binary systems, we find that in these low-metallicity high-density environments most bright XRBs should be observed near but outside their parent cluster, most having been ejected.

Analyzing the orbital period and eccentricity distributions of BH binary systems we find that BH binaries (BH--BH systems and more massive BH--MS systems) surviving in a dense cluster after $\sim 100\,\rm Myr$, have $P_{\rm orb}<10^3\,\rm d$ 
and $e>0.2$. Only low-mass ($M_{\rm BH}<10\,M_{\odot}$) 
BH--MS systems residing away from the core can have longer orbital periods.

We also find 
at very early times ($<5\,\rm Myr$) some collisional mergers (unlike the CE mergers mentioned in B04 and B06) of massive stars. Primordial binaries with
massive components have the largest collision cross section and will contribute the most to these collisions. These early collisional mergers either increase 
the mass of the primary in the binary system, or form a massive single star (when all the participating stars merge together), eventually forming a more 
massive BH. 
 
Exchange interactions are rare during the initial phase of the cluster evolution ($<10\,\rm Myr$), and become noticeable only at later times. In our simulations as the cluster evolves beyond $10\,\rm Myr$, when dynamics becomes important, 
the most massive stars  have already undergone SNe. Hence  exchange interactions mainly happen among massive BH binaries, massive single BHs, and 
intermediate-mass MS stars which still exist in the cluster at that time. In most of these interactions a low-mass MS companion is exchanged for another more massive BH. 
A high primordial binary fraction (with all other input parameters remaining constant) leads to a higher rate of exchange interactions when dynamics is important (dense clusters older than $\sim 10\,\rm Myr$). However, a large fraction of  primordial binaries increases the average particle mass in the cluster (with all other initial parameters remaining same) and hence the dynamical friction timescale. 
Thus, even with a very high primordial binary fraction for massive stars, as in our model W1, and a correspondingly higher rate of exchange interactions, we do not find a significant increase in BH--MS pairs, when compared to our model W2.  

As expected, more massive BHs are formed at lower metallicities in our simulations. These more massive BHs concentrate in the cluster core through mass 
segregation and, in most of our models, $\sim60\%$ acquire a binary companion. The timescale for BHs to mass segregate and acquire a binary companion is 
always more than $\sim 10\,\rm Myr$ and the binary companion is almost always another BH. Hence, it is not clear whether the formation rate of X-ray binaries is enhanced
by exchange interactions in young massive clusters.

Comparing our population synthesis results (for low-density cluster models) with those of B04 and B06, we find that our stellar evolution prescriptions (BSE) agree 
reasonably with those of the StarTrack code. Our study with full dynamics and stellar evolution shows that dynamics play a major role in determining the numbers and properties of BHs in young star clusters. Dynamical interactions play an important role in: (i) ejecting more and more BHs from the cluster, 
(ii) increasing the eccentricity of retained binary population, 
(iii) decreasing the orbital period of the retained binary population, (iv) increasing the probability of having a comparatively more massive companion through exchange interactions and, (v) forming massive BHs through early collisional mergers. 
All these effects can significantly modify the properties of BHs in binaries, including X-ray properties. Hence, we conclude 
that population synthesis studies are not adequate for analyzing dense young stellar clusters.

We thank Meagan Morscher and Stefan Umbreit for useful discussions regarding the evolution of dense star clusters. We also express our thanks to the anonymous referee and to Vicky Kalogera for many valuable comments. This work was supported by NASA ATP   
Grant NNX09A036G and NSF Grant PHY-0855592 at Northwestern University.  F. A. Rasio also acknowledges support from NSF Grant PHY-1066293 at the Aspen Center for Physics.

\begin{deluxetable}{cccccccc}  
  \tablecolumns{8} 
  \tablewidth{0pc}
  \tablecaption{SUMMARY OF ALL THE SIMULATIONS WITH METALLICITY $Z=0.001$\label{t1}  }

\tablehead{   
 \colhead{Name}&
  \colhead{$N$} &
  \colhead{$f_{b}$} &
  \colhead{$r_{\rm v}$} &
   \colhead{$\rho_c(0)$} & 
  \colhead{$t_{\rm rh}(0)$} & 
  \colhead{dynamics} & 
  \colhead{$f_{b\rm (ej)}$ at $100\,\rm Myr$}\\
  \colhead{} &
  \colhead{} &
  \colhead{($\%$)} &
  \colhead{(pc)} &
   \colhead{($10^5 M_\odot\,{\rm pc}^{-3}$)} &
  \colhead{($10^8$ years)} &
  \colhead{(y/n)} &
  \colhead{($\sim\%$)}
}
\startdata

\multicolumn{8}{c}{Orbital period distribution: \citet{Iva05} } \\

\hline
\multicolumn{7}{c}{Standard Model } \\

A & 5x$10^{5}$ & 50 & 1.25 & 1.3 &  2.0 & y & 57\\
B & 5x$10^{5}$ & 50 &  & &  & n & 17\\
\hline

\hline
A1 & 2x$10^{5}$ &  0 & 1.25& 0.26  & 2.0 & y & 0 \\
B1 & 2x$10^{5}$ &  0 & 1.74 & 0.10& 3.3 & y & 0\\
C1 & 2x$10^{5}$ &  0 & 3 & 0.02 & 7.5  & y & 0\\
D1 & 2x$10^{5}$ &  0 &  & &   & n  & 0\\

A2 & 2x$10^{5}$ & 10 & 1.25 & 0.26 & 2.0 & y & 35\\
B2 & 2x$10^{5}$ & 10 & 1.74 & 0.10 & 3.2 & y & 20 \\
C2 & 2x$10^{5}$ & 10 & 3 & 0.02 & 7.3 & y & 15\\
D2 & 2x$10^{5}$ & 10 & &    &  & n & 15\\

A3 & 2x$10^{5}$ & 30 & 1.25 & 0.32& 1.8 & y & 33 \\
B3 & 2x$10^{5}$ & 30 & 1.74 & 0.12 & 3.1 & y & 25 \\
C3 & 2x$10^{5}$ & 30 & 3 & 0.02 & 7.0 & y & 15\\
D3 & 2x$10^{5}$ & 30 &  &   &  & n & 15\\

A4 & 2x$10^{5}$ & 50 & 1.1& 0.32 & 1.8 & y & 32\\
B4 & 2x$10^{5}$ & 50 & 1.74 & 0.12 &  2.9 & y & 20 \\
C4 & 2x$10^{5}$ & 50 & 3 & 0.02 & 6.7 & y & 15\\
D4 & 2x$10^{5}$ & 50 &  &   &  & n & 15\\

\hline

I1 & 4x$10^{5}$ & 0 & 1.25& 0.52 & 2.6 &  y &  0\\
J1 & 4x$10^{5}$ & 0 & 1.64 &0.23 & 3.9 & y & 0 \\
K1 & 4x$10^{5}$ & 0 & 3 &0.04 & 9.7 & y & 0 \\
L1 & 4x$10^{5}$ & 0 &  & &   & n &  0\\

I2 & 4x$10^{5}$ & 10 & 1.25 &0.52 & 2.6 &  y & 43\\
J2 & 4x$10^{5}$ & 10 & 1.64 & 0.24 & 3.8 & y & 30\\
K2 & 4x$10^{5}$ & 10 & 3 & 0.04 & 9.4 &  y & 23\\
L2 & 4x$10^{5}$ & 10 &   & &   & n & 15\\

I3 & 4x$10^{5}$ & 30 & 1.25& 0.61 & 2.4 & y & 45\\
J3 & 4x$10^{5}$ & 30 & 1.64& 0.26 & 3.7 &   y & 34\\
K3 & 4x$10^{5}$ & 30 & 3 &0.04 &  9.0 & y & 20\\
L3 & 4x$10^{5}$ & 30 &   &  &  & n & 15 \\

I4 & 4x$10^{5}$ & 50 & 1.25 & 0.66&  2.3 & y & 45\\
J4 & 4x$10^{5}$ & 50 & 1.40 & 0.30 & 3.4 & y & 32\\
K4 & 4x$10^{5}$ & 50 & 1.64 & 0.49 & 8.5 & y & 22\\
L4 & 4x$10^{5}$ & 50 &   &  &  & n & 10\\

\hline
M1 & 5x$10^{5}$ & 0 & 1.25& 0.98 & 2.4 &  y & 0 \\
N1 & 5x$10^{5}$ & 0 & 1.40 & 0.38 & 3.7  & y & 0 \\
O1 & 5x$10^{5}$ & 0 & 1.64 & 0.24 & 4.7 & y & 0\\
P1 & 5x$10^{5}$ & 0 &  & &   &  n &  0 \\

M2 & 5x$10^{5}$ & 40 & 1.25 &1.13& 2.14 & y & 56 \\
N2 & 5x$10^{5}$ & 40 & 1.40 &0.46 &  3.34 & y & 40\\
O2 & 5x$10^{5}$ & 40 & 1.64 & 0.30 & 4.22 & y & 13 \\
P2 & 5x$10^{5}$ & 40 &   & &  & n & 10\\

N4 & 5x$10^{5}$ & 50 & 1.40 &0.45 & 3.37 &  y & 36\\
O4 & 5x$10^{5}$ & 50 & 1.64 & 0.20 & 4.2 &  y & 22\\

\hline

\enddata

\tablecomments{  Here $N$, $r_{ v}$, $\rho_c(0)$ and $t_{\rm rh}(0)$ are the initial number of stars, the initial virial radius, the initial core density and the initial half mass relaxation time respectively. $f_{b}$ denotes the primordial binary fraction in the cluster while $f_{b\rm (ej)}$ denotes the fraction of all BHs in binaries that got ejected from the cluster. All initial models start with a Kroupa IMF (mass range $0.08-150\, M_{\odot}$). All the simulations have a \citet{Arz02} kick distribution and \citet{Hur02} wind prescription. The maximum neutron star mass in these simulations is, $M_{\rm max,NS}=3\, M_{\odot}$. The prescription for the masses of the compact remnants are as implemented in B06.
The percentage values ($f_{b\rm (ej)}$) are calculated after averaging the binary BH ejection fraction over 5 simulation results.}

\end{deluxetable}

\begin{deluxetable}{cccccccccc}  
  \tablecolumns{8} 
  \tablewidth{0pc}
  \tablecaption{SUMMARY OF ALL THE SIMULATIONS WITH METALLICITIES $Z=0.0002$ AND $Z=0.02$ \label{t2}  }

\tablehead{   
 \colhead{Name}&
  \colhead{$N$} &
  \colhead{$f_{b}$} &
  \colhead{$r_{\rm v}$} &
     \colhead{$\rho_c(0)$} & 
  \colhead{$t_{\rm rh}(0)$} &
 
  \colhead{dynamics} & 
  \colhead{$f_{b\rm (ej)}$ at $100\,\rm Myr$}\\
  \colhead{} &
  \colhead{} &
  \colhead{($\%$)} &
  \colhead{(pc)} &
   \colhead{($10^5 M_\odot\,{\rm pc}^{-3}$)} &
 \colhead{($10^8$ years)} &
  \colhead{(y/n)} &
  \colhead{($\sim\%$)}
}
\startdata

\multicolumn{8}{c}{Orbital period distribution: \citet{Hur05} } \\
\multicolumn{8}{c}{Metallicity: $Z=0.0002$ } \\

\hline
AA1 & 5x$10^{5}$ &  0 & 1.25 & 1.30  & 2.0 & y & 0\\
BB1 & 5x$10^{5}$ & 0  &  1.72 & 0.26 & 4.5 & y  & 0\\
CC1 & 5x$10^{5}$ & 0  & 3 & 0.05 &  &y & 0\\
DD1 & 5x$10^{5}$ &  0 &  & &   &n  & 0\\

AA2 & 5x$10^{5}$ & 30 & 1.25 & 1.07 & 2.2 & y & 62\\
BB2 & 5x$10^{5}$ & 30 & 1.72 & 0.26 & 4.5 & y & 45\\
CC2 & 5x$10^{5}$ & 30 & 3 & 0.05 & 10.4 & y & 11\\
DD2 & 5x$10^{5}$ & 30 &  &    &  &n & 11\\

AA3 & 5x$10^{5}$ & 40 & 1.25 & 1.61 & 1.8 & y & 60\\
BB3 & 5x$10^{5}$ & 40 & 1.72 & 0.31 & 4.1 & y & 40\\
CC3 & 5x$10^{5}$ & 40 & 3 & 0.06 & 9.3 & y & 11\\
DD3 & 5x$10^{5}$ & 40 &  &  &  &n & 11\\

AA4 & 5x$10^{5}$ & 50 & 1.25 & 1.61& 1.8 & y & 60\\
BB4 & 5x$10^{5}$ & 50 & 1.72 & 0.30 & 4.1 & y & 40\\
CC4 & 5x$10^{5}$ & 50 & 3 & 0.06 & 9.3 & y & 10\\
DD4 & 5x$10^{5}$ & 50 &  &  &  &n & 10\\
\hline
\multicolumn{7}{c}{Metallicity : $Z=0.02$ } \\
\hline
EE1 & 5x$10^{5}$ &  0 & 1.25& 1.07  & 2.0 & y & 0\\
FF1 & 5x$10^{5}$ & 0  &  1.47& 0.42 & 3.6 & y  &0\\
GG1 & 5x$10^{5}$ & 0  &  1.72& 0.26  & 4.5 & y & 0 \\
HH1 & 5x$10^{5}$ &  0 & &  &  & n  &0\\

EE2 & 5x$10^{5}$ & 30 & 1.25 & 1.30 & 2.2 & y & 15\\
FF2 & 5x$10^{5}$ & 30 & 1.47 & 0.40 & 3.5 & y & 10\\
GG2 & 5x$10^{5}$ & 30 & 1.72 &0.26 & 4.5 & y & 10\\
HH2 & 5x$10^{5}$ & 30 &  &   &  & n & 10\\

EE3 & 5x$10^{5}$ & 40 & 1.25& 1.61 & 1.8 &  y & 16\\
FF3 & 5x$10^{5}$ & 40 & 1.47 & 0.47& 3.3 & y & 11\\
GG3 & 5x$10^{5}$ & 40 & 1.72& 0.31 & 4.1 &y & 10 \\
HH3 & 5x$10^{5}$ & 40 &  &    &  & n & 10 \\

EE4 & 5x$10^{5}$ & 50 & 1.25 & 1.61 &  1.8 & y & 16\\
FF4 & 5x$10^{5}$ & 50 & 1.47 &  0.51 & 3.2 & y & 15\\
GG4 & 5x$10^{5}$ & 50 & 1.72 & 0.31 & 4.1 & y & 10 \\
HH4 & 5x$10^{5}$ & 50 &   &  &  &n & 6\\\hline

\enddata
\tablecomments{  Same as Table \ref{t1} except that all the simulations have a Maxwellian kick distribution with $\sigma=265\,\rm Km\, s^{-1}$ and wind prescription as well as remnant masses of NSs and BHs as implemented in \citet{Bel10}. The maximum neutron star mass in these simulations is $M_{\rm max,NS}=2.5\, M_{\odot}$.}

\end{deluxetable}


\begin{deluxetable}{cccccccc}  
  \tablecolumns{8} 
  \tablewidth{0pc}
  \tablecaption{SIMULATIONS WITH BINARY FRACTION AS IN WESTERLUND 1\label{t3}  }
  
\tablehead{   
 \colhead{Name}&
  \colhead{$N$} &
  \colhead{$f_{b}$} &
  \colhead{$r_{\rm v}$} &
    \colhead{$\rho_c(0)$} &
  \colhead{$t_{\rm rh}(0)$} &
  \colhead{dynamics} & 
  \colhead{$f_{b\rm (ej)}$ at $100\,\rm Myr$}\\
  \colhead{} &
  \colhead{} &
  \colhead{($\%$)} &
  \colhead{(pc)} &
   \colhead{($10^5 M_\odot\,{\rm pc}^{-3}$)} &
  \colhead{($10^8$ years)} &
  \colhead{(y/n)} &
  \colhead{($\%$)}
}
\startdata
\multicolumn{8}{c}{Metallicity : $Z=0.001$ } \\
\multicolumn{8}{c}{Orbital period distribution : \citet{Hur05} } \\

\hline
W1 & 5x$10^{5}$ & 50 & 1.25& 1.61 & 1.8 & y & 63.1\\
W2 & 5x$10^{5}$ & 50 &  &  & &n & 15.9\\
\\
\hline
\enddata
\tablecomments{ Same as Table \ref{t1} except that all the simulations use a Maxwellian kick velocity distribution with $\sigma=150\,\rm km\, s^{-1}$ and wind prescription as well as remnant masses of NSs and BHs as implemented in \citet{Bel10}.}

\end{deluxetable}


\begin{deluxetable}{cccccccc}  
  \tablecolumns{7} 
  \tablewidth{0pc}
  \tablecaption{BLACK HOLE POPULATIONS RETAINED IN / EJECTED FROM STANDARD MODEL B \label{t4} }

\tablehead{   
 \colhead{Type}&
  \colhead{$10\, \rm Myr$} &
  \colhead{$20\, \rm Myr$} &
  \colhead{$40\, \rm Myr$} &
  \colhead{$60\, \rm Myr$} &
  \colhead{$80\, \rm Myr$} &
 \colhead{$100\, \rm Myr$} \\
  \colhead{} &
\colhead{($0.05\,t_{\rm rh}(0)$)} &
  \colhead{($0.10\,t_{\rm rh}(0)$)} &
  \colhead{($0.20\,t_{\rm rh}(0)$)} &
  \colhead{($0.29\,t_{\rm rh}(0)$)} &
  \colhead{($0.39\,t_{\rm rh}(0)$)} &
   \colhead{($0.49\,t_{\rm rh}(0)$)}
}
\startdata

BH--MS  & 34/7 & 28/9 & 19/9 & 15/9  & 14/9  & 13/9 \\
BH--HG  & 0 & 0 & 0 & 0 & 0 & 0 \\
BH--RG  & 0 & 0 & 0 & 0 & 0 & 0 \\
BH--CheB  & 4/0 & 1/0 & 0 & 1/0 & 1/0 & 1/0 \\
BH---AGB  & 0 & 1/0 & 0 & 0 & 0 & 0 \\
BH--He  & 1/0 & 0 & 0 & 0 & 0 & 0 \\
BH--WD & 0 & 0 & 0 & 0 & 0 & 0 \\
BH--NS & 0 & 0 & 1/0 & 1/0 & 1/0 & 1/0 \\
BH--BH & 34/2 & 38/3 & 38/3 & 38/3  & 38/3  & 38/3 \\

Total in Binaries & 73/9 & 68/12 & 58/12 & 55/12  & 55/12  & 55/12 \\




Total Singles & 171/0 & 218/10 & 230/10 & 230/10  & 230/10  & 230/10 \\

\enddata
\tablecomments{  Here $t_{\rm rh}(0)$ is the initial half-mass relaxation time.}
\end{deluxetable}

\begin{deluxetable}{cccccccc}  
  \tablecolumns{7} 
  \tablewidth{0pc}
  \tablecaption{BLACK HOLE POPULATIONS RETAINED IN / EJECTED FROM STANDARD MODEL A \label{t5}  }

\tablehead{   
 \colhead{Type}&
  \colhead{$10\, \rm Myr$} &
  \colhead{$20\, \rm Myr$} &
  \colhead{$40\, \rm Myr$} &
  \colhead{$60\, \rm Myr$} &
  \colhead{$80\, \rm Myr$} &
 \colhead{$100\, \rm Myr$} \\
  \colhead{} &
\colhead{($0.05\,t_{\rm rh}(0)$)} &
  \colhead{($0.10\,t_{\rm rh}(0)$)} &
  \colhead{($0.20\,t_{\rm rh}(0)$)} &
  \colhead{($0.29\,t_{\rm rh}(0)$)} &
  \colhead{($0.39\,t_{\rm rh}(0)$)} &
   \colhead{($0.49\,t_{\rm rh}(0)$)}
}
\startdata

BH--MS  & 11/5 & 10/8 & 6/8 & 5/8  & 5/8  & 9/8 \\
BH--HG  & 0 & 0 & 0 & 0 & 0 & 0 \\
BH--RG  & 0 & 0 & 0 & 0 & 0 & 0 \\
BH--CheB  & 0 & 1/0 & 0 & 0 & 0 & 0 \\
BH--AGB  & 0 & 0 & 0 & 0 & 0 & 0 \\
BH--He  & 0 & 0 & 0 & 0 & 0 & 0 \\
BH--WD & 0 & 0 & 0 & 0 & 0 & 0 \\
BH--NS & 0 & 0 & 0 & 0 & 0 & 0 \\
BH--BH & 5/3 & 3/3 & 3/5 & 6/8  & 3/11  & 5/12 \\

Total in Binaries & 16/8 & 14/11 & 9/13 & 11/16  & 8/19  & 15/20 \\




Total Singles & 226/2 & 280/8 & 270/22 & 241/40  & 233/49  & 214/57 \\

\enddata
\tablecomments{ Here $t_{\rm rh}(0)$ is the initial half-mass relaxation time.}

\end{deluxetable}


\bibliographystyle{apj}
\bibliography{ymc}

\begin{thebibliography}{84}
\expandafter\ifx\csname natexlab\endcsname\relax\def\natexlab#1{#1}\fi

\bibitem[{{Aasi} {et~al.}(2013)}]{Aas13}
{Aasi}, J., {Abadie}, J., {Abbott}, B.~P., {Abbott}, R., {Abbott}, T.~D.,
  {Abernathy}, M., {Accadia}, T., {Acernese}, F., {Adams}, C., {Adams}, T., \&
  et~al. 2013, \prd, 87, 022002

\bibitem[{{Arzoumanian} {et~al.}(2002){Arzoumanian}, {Chernoff}, \&
  {Cordes}}]{Arz02}
{Arzoumanian}, Z., {Chernoff}, D.~F., \& {Cordes}, J.~M. 2002, \apj, 568, 289

\bibitem[{{Belczynski} {et~al.}(2010)}]{Bel10}
{Belczynski}, K., {Bulik}, T., {Fryer}, C.~L., {Ruiter}, A., {Valsecchi}, F.,
  {Vink}, J.~S., \& {Hurley}, J.~R. 2010, \apj, 714, 1217

\bibitem[{{Belczynski} {et~al.}(2008)}]{Bel08}
{Belczynski}, K., {Kalogera}, V., {Rasio}, F.~A., {Taam}, R.~E., {Zezas}, A.,
  {Bulik}, T., {Maccarone}, T.~J., \& {Ivanova}, N. 2008, \apjs, 174, 223

\bibitem[{{Belczynski} {et~al.}(2004){Belczynski}, {Sadowski}, \&
  {Rasio}}]{Bel04}
{Belczynski}, K., {Sadowski}, A., \& {Rasio}, F.~A. 2004, \apj, 611, 1068

\bibitem[{{Belczynski} {et~al.}(2006)}]{Bel06}
{Belczynski}, K., {Sadowski}, A., {Rasio}, F.~A., \& {Bulik}, T. 2006, \apj,
  650, 303

\bibitem[{{Bellazzini} {et~al.}(1995)}]{Bel95}
{Bellazzini}, M., {Pasquali}, A., {Federici}, L., {Ferraro}, F.~R., \& {Pecci},
  F.~F. 1995, \apj, 439, 687

\bibitem[{{Blecha} {et~al.}(2006)}]{Ble06}
{Blecha}, L., {Ivanova}, N., {Kalogera}, V., {Belczynski}, K., {Fregeau}, J.,
  \& {Rasio}, F. 2006, \apj, 642, 427

\bibitem[{{Breen} \& {Heggie}(2013)}]{Bre13}
{Breen}, P.~G. \& {Heggie}, D.~C. 2013, \mnras, 432, 2779

\bibitem[{{Clark} {et~al.}(2008)}]{Cla08}
{Clark}, J.~S., {Muno}, M.~P., {Negueruela}, I., {Dougherty}, S.~M.,
  {Crowther}, P.~A., {Goodwin}, S.~P., \& {de Grijs}, R. 2008, \aap, 477, 147

\bibitem[{{Clark} {et~al.}(2005)}]{Cla05}
{Clark}, J.~S., {Negueruela}, I., {Crowther}, P.~A., \& {Goodwin}, S.~P. 2005,
  \aap, 434, 949

\bibitem[{{Clark} {et~al.}(2011)}]{Cla11}
{Clark}, J.~S., {Ritchie}, B.~W., {Negueruela}, I., {Crowther}, P.~A.,
  {Damineli}, A., {Jablonski}, F.~J., \& {Langer}, N. 2011, \aap, 531, A28

\bibitem[{{Colbert} \& {Mushotzky}(1999)}]{Col99}
{Colbert}, E.~J.~M. \& {Mushotzky}, R.~F. 1999, Advances in Space Research, 23,
  847

\bibitem[{{Colpi} {et~al.}(2003){Colpi}, {Mapelli}, \& {Possenti}}]{Col03}
{Colpi}, M., {Mapelli}, M., \& {Possenti}, A. 2003, \apj, 599, 1260

\bibitem[{{Copperwheat} {et~al.}(2007)}]{Cop07}
{Copperwheat}, C., {Cropper}, M., {Soria}, R., \& {Wu}, K. 2007, \mnras, 376,
  1407

\bibitem[{{Cowley}(1994)}]{Cow94}
{Cowley}, A.~P. 1994, in Astronomical Society of the Pacific Conference Series,
  Vol.~56, Interacting Binary Stars, ed. A.~W. {Shafter}, 160

\bibitem[{{Crowther} {et~al.}(2006)}]{Cro06}
{Crowther}, P.~A., {Hadfield}, L.~J., {Clark}, J.~S., {Negueruela}, I., \&
  {Vacca}, W.~D. 2006, \mnras, 372, 1407

\bibitem[{{Elmegreen} \& {Efremov}(1997)}]{Elm97}
{Elmegreen}, B.~G. \& {Efremov}, Y.~N. 1997, \apj, 480, 235

\bibitem[{{Fabbiano} \& {White}(2006)}]{Fab06}
{Fabbiano}, G. \& {White}, N.~E. 2006, {Compact stellar X-ray sources in normal
  galaxies}, ed. W.~H.~G. {Lewin} \& M.~{van der Klis}, 475--506

\bibitem[{{Fabbiano} {et~al.}(2001){Fabbiano}, {Zezas}, \& {Murray}}]{Fab01}
{Fabbiano}, G., {Zezas}, A., \& {Murray}, S.~S. 2001, \apj, 554, 1035

\bibitem[{{Fragos} {et~al.}(2009)}]{Fra09}
{Fragos}, T., {Willems}, B., {Kalogera}, V., {Ivanova}, N., {Rockefeller}, G.,
  {Fryer}, C.~L., \& {Young}, P.~A. 2009, \apj, 697, 1057

\bibitem[{{Fregeau} {et~al.}(2004)}]{Fre04}
{Fregeau}, J.~M., {Cheung}, P., {Portegies Zwart}, S.~F., \& {Rasio}, F.~A.
  2004, \mnras, 352, 1

\bibitem[{{Fregeau} {et~al.}(2002)}]{Fre02}
{Fregeau}, J.~M., {Joshi}, K.~J., {Portegies Zwart}, S.~F., \& {Rasio}, F.~A.
  2002, \apj, 570, 171

\bibitem[{{Fregeau} \& {Rasio}(2007)}]{Fre07}
{Fregeau}, J.~M. \& {Rasio}, F.~A. 2007, \apj, 658, 1047

\bibitem[{{Freitag} {et~al.}(2006){Freitag}, {Rasio}, \& {Baumgardt}}]{Fre06}
{Freitag}, M., {Rasio}, F.~A., \& {Baumgardt}, H. 2006, \mnras, 368, 121

\bibitem[{{Fryer}(1999)}]{Fry99}
{Fryer}, C.~L. 1999, \apj, 522, 413

\bibitem[{{Gon{\c c}alves} \& {Soria}(2006)}]{Gon06}
{Gon{\c c}alves}, A.~C. \& {Soria}, R. 2006, \mnras, 371, 673

\bibitem[{{Goswami} {et~al.}(2012)}]{San12}
{Goswami}, S., {Umbreit}, S., {Bierbaum}, M., \& {Rasio}, F.~A. 2012, \apj,
  752, 43

\bibitem[{{Gualandris} {et~al.}(2005)}]{Gua05}
{Gualandris}, A., {Colpi}, M., {Portegies Zwart}, S., \& {Possenti}, A. 2005,
  \apj, 618, 845

\bibitem[{{G{\"u}rkan} {et~al.}(2004){G{\"u}rkan}, {Freitag}, \&
  {Rasio}}]{Gur04}
{G{\"u}rkan}, M.~A., {Freitag}, M., \& {Rasio}, F.~A. 2004, \apj, 604, 632

\bibitem[{{Heger} {et~al.}(2003)}]{Heg03}
{Heger}, A., {Fryer}, C.~L., {Woosley}, S.~E., {Langer}, N., \& {Hartmann},
  D.~H. 2003, \apj, 591, 288

\bibitem[{{Heger} \& {Woosley}(2002)}]{Heg02}
{Heger}, A. \& {Woosley}, S.~E. 2002, \apj, 567, 532

\bibitem[{{Heggie}(1975)}]{Heg75}
{Heggie}, D.~C. 1975, \mnras, 173, 729

\bibitem[{{H{\'e}non}(1971)}]{Henon}
{H{\'e}non}, M.~H. 1971, \apss, 14, 151

\bibitem[{{Hobbs} {et~al.}(2005)}]{Hob05}
{Hobbs}, G., {Lorimer}, D.~R., {Lyne}, A.~G., \& {Kramer}, M. 2005, \mnras,
  360, 974

\bibitem[{{Hurley} {et~al.}(2005)}]{Hur05}
{Hurley}, J.~R., {Pols}, O.~R., {Aarseth}, S.~J., \& {Tout}, C.~A. 2005,
  \mnras, 363, 293

\bibitem[{{Hurley} {et~al.}(2000){Hurley}, {Pols}, \& {Tout}}]{Hur00}
{Hurley}, J.~R., {Pols}, O.~R., \& {Tout}, C.~A. 2000, \mnras, 315, 543

\bibitem[{{Hurley} {et~al.}(2002){Hurley}, {Tout}, \& {Pols}}]{Hur02}
{Hurley}, J.~R., {Tout}, C.~A., \& {Pols}, O.~R. 2002, \mnras, 329, 897

\bibitem[{{Iben} {et~al.}(1997){Iben}, {Tutukov}, \& {Fedorova}}]{Ibe97}
{Iben}, Jr., I., {Tutukov}, A.~V., \& {Fedorova}, A.~V. 1997, \apj, 486, 955

\bibitem[{{Ivanova} {et~al.}(2005)}]{Iva05}
{Ivanova}, N., {Belczynski}, K., {Fregeau}, J.~M., \& {Rasio}, F.~A. 2005,
  \mnras, 358, 572

\bibitem[{{Janka}(2013)}]{Jan13}
{Janka}, H.-T. 2013, \mnras, 434, 1355

\bibitem[{{Kaaret} {et~al.}(2004)}]{Kaa04}
{Kaaret}, P., {Alonso-Herrero}, A., {Gallagher}, J.~S., {Fabbiano}, G.,
  {Zezas}, A., \& {Rieke}, M.~J. 2004, \mnras, 348, L28

\bibitem[{{Kaaret} {et~al.}(2001)}]{Kaa01}
{Kaaret}, P., {Prestwich}, A.~H., {Zezas}, A., {Murray}, S.~S., {Kim}, D.-W.,
  {Kilgard}, R.~E., {Schlegel}, E.~M., \& {Ward}, M.~J. 2001, \mnras, 321, L29

\bibitem[{{Kiel} \& {Hurley}(2009)}]{Kie09}
{Kiel}, P.~D. \& {Hurley}, J.~R. 2009, \mnras, 395, 2326

\bibitem[{{King} {et~al.}(2001)}]{Kin01}
{King}, A.~R., {Davies}, M.~B., {Ward}, M.~J., {Fabbiano}, G., \& {Elvis}, M.
  2001, \apjl, 552, L109

\bibitem[{{Kulkarni} {et~al.}(1993){Kulkarni}, {Hut}, \& {McMillan}}]{Kul93}
{Kulkarni}, S.~R., {Hut}, P., \& {McMillan}, S. 1993, \nat, 364, 421

\bibitem[{{Kundu} {et~al.}(2002){Kundu}, {Maccarone}, \& {Zepf}}]{Kun02}
{Kundu}, A., {Maccarone}, T.~J., \& {Zepf}, S.~E. 2002, \apjl, 574, L5

\bibitem[{{Lattimer} \& {Yahil}(1989)}]{Lat89}
{Lattimer}, J.~M. \& {Yahil}, A. 1989, \apj, 340, 426

\bibitem[{{Linden} {et~al.}(2010)}]{Lin10}
{Linden}, T., {Kalogera}, V., {Sepinsky}, J.~F., {Prestwich}, A., {Zezas}, A.,
  \& {Gallagher}, J.~S. 2010, \apj, 725, 1984

\bibitem[{{Lyne} \& {Lorimer}(1994)}]{Lyn94}
{Lyne}, A.~G. \& {Lorimer}, D.~R. 1994, \nat, 369, 127

\bibitem[{{Mackey} {et~al.}(2007)}]{Mac07}
{Mackey}, A.~D., {Wilkinson}, M.~I., {Davies}, M.~B., \& {Gilmore}, G.~F. 2007,
  \mnras, 379, L40

\bibitem[{{Mapelli} {et~al.}(2009){Mapelli}, {Colpi}, \& {Zampieri}}]{Map09}
{Mapelli}, M., {Colpi}, M., \& {Zampieri}, L. 2009, \mnras, 395, L71

\bibitem[{{Mapelli} {et~al.}(2011)}]{Map11}
{Mapelli}, M., {Ripamonti}, E., {Zampieri}, L., \& {Colpi}, M. 2011, ArXiv
  e-prints

\bibitem[{{Mapelli} {et~al.}(2010)}]{Map10}
{Mapelli}, M., {Ripamonti}, E., {Zampieri}, L., {Colpi}, M., \& {Bressan}, A.
  2010, \mnras, 408, 234

\bibitem[{{Mapelli} {et~al.}(2013)}]{Map13}
{Mapelli}, M., {Zampieri}, L., {Ripamonti}, E., \& {Bressan}, A. 2013, \mnras,
  429, 2298

\bibitem[{{Miller} {et~al.}(2003)}]{Mil03}
{Miller}, J.~M., {Fabbiano}, G., {Miller}, M.~C., \& {Fabian}, A.~C. 2003,
  \apjl, 585, L37

\bibitem[{{Miller} {et~al.}(2004){Miller}, {Fabian}, \& {Miller}}]{Mil04}
{Miller}, J.~M., {Fabian}, A.~C., \& {Miller}, M.~C. 2004, \apj, 607, 931

\bibitem[{{Mitsuda} {et~al.}(1984)}]{Mit84}
{Mitsuda}, K., {Inoue}, H., {Koyama}, K., {Makishima}, K., {Matsuoka}, M.,
  {Ogawara}, Y., {Suzuki}, K., {Tanaka}, Y., {Shibazaki}, N., \& {Hirano}, T.
  1984, \pasj, 36, 741

\bibitem[{{Morscher} {et~al.}(2013)}]{Mor13}
{Morscher}, M., {Umbreit}, S., {Farr}, W.~M., \& {Rasio}, F.~A. 2013, \apjl,
  763, L15

\bibitem[{{Mushotzky}(2000)}]{Mus00}
{Mushotzky}, R.~F. 2000, in Bulletin of the American Astronomical Society,
  Vol.~32, American Astronomical Society Meeting Abstracts \#196, 726

\bibitem[{{Negueruela} \& {Clark}(2005)}]{Neg05}
{Negueruela}, I. \& {Clark}, J.~S. 2005, \aap, 436, 541

\bibitem[{{Pakull} \& {Mirioni}(2001)}]{Pak01}
{Pakull}, M.~W. \& {Mirioni}, L. 2001, in Astronomische Gesellschaft Meeting
  Abstracts, Vol.~18, Astronomische Gesellschaft Meeting Abstracts, ed.
  {E.~R.~Schielicke}, 112

\bibitem[{{Pakull} \& {Mirioni}(2002)}]{Pak02}
{Pakull}, M.~W. \& {Mirioni}, L. 2002, ArXiv Astrophysics e-prints

\bibitem[{{Patruno} {et~al.}(2005)}]{Pat05}
{Patruno}, A., {Colpi}, M., {Faulkner}, A., \& {Possenti}, A. 2005, \mnras,
  364, 344

\bibitem[{{Portegies Zwart} {et~al.}(2004)}]{Por04}
{Portegies Zwart}, S.~F., {Baumgardt}, H., {Hut}, P., {Makino}, J., \&
  {McMillan}, S.~L.~W. 2004, \nat, 428, 724

\bibitem[{{Portegies Zwart} \& {McMillan}(2000)}]{Por00}
{Portegies Zwart}, S.~F. \& {McMillan}, S.~L.~W. 2000, \apjl, 528, L17

\bibitem[{{Portegies Zwart} {et~al.}(2010){Portegies Zwart}, {McMillan}, \&
  {Gieles}}]{Zwa10}
{Portegies Zwart}, S.~F., {McMillan}, S.~L.~W., \& {Gieles}, M. 2010, \araa,
  48, 431

\bibitem[{{Portegies Zwart} \& {Verbunt}(1996)}]{Por96}
{Portegies Zwart}, S.~F. \& {Verbunt}, F. 1996, \aap, 309, 179

\bibitem[{{Prestwich} {et~al.}(2013)}]{Pre13}
{Prestwich}, A.~H., {Tsantaki}, M., {Zezas}, A., {Jackson}, F.~E., {Roberts},
  T.~P., {Foltz}, R., {Linden}, T., \& {Kalogera}, V. 2013, ArXiv e-prints

\bibitem[{{Rappaport} {et~al.}(2005){Rappaport}, {Podsiadlowski}, \&
  {Pfahl}}]{Rap05}
{Rappaport}, S.~A., {Podsiadlowski}, P., \& {Pfahl}, E. 2005, \mnras, 356, 401

\bibitem[{{Rasio} \& {Heggie}(1995)}]{Ras95}
{Rasio}, F.~A. \& {Heggie}, D.~C. 1995, \apjl, 445, L133

\bibitem[{{Sepinsky} {et~al.}(2005){Sepinsky}, {Kalogera}, \&
  {Belczynski}}]{Sep04}
{Sepinsky}, J., {Kalogera}, V., \& {Belczynski}, K. 2005, \apjl, 621, L37

\bibitem[{{Sigurdsson} \& {Phinney}(1993)}]{Sig93}
{Sigurdsson}, S. \& {Phinney}, E.~S. 1993, \apj, 415, 631

\bibitem[{{Skinner} {et~al.}(2006)}]{Ski06}
{Skinner}, S.~L., {Simmons}, A.~E., {Zhekov}, S.~A., {Teodoro}, M., {Damineli},
  A., \& {Palla}, F. 2006, \apjl, 639, L35

\bibitem[{{Soria} {et~al.}(2005)}]{Sor05}
{Soria}, R., {Cropper}, M., {Pakull}, M., {Mushotzky}, R., \& {Wu}, K. 2005,
  \mnras, 356, 12

\bibitem[{{Stobbart} {et~al.}(2006){Stobbart}, {Roberts}, \& {Wilms}}]{Ro06}
{Stobbart}, A.-M., {Roberts}, T.~P., \& {Wilms}, J. 2006, \mnras, 368, 397

\bibitem[{{Swartz} {et~al.}(2008){Swartz}, {Soria}, \& {Tennant}}]{Swa08}
{Swartz}, D.~A., {Soria}, R., \& {Tennant}, A.~F. 2008, \apj, 684, 282

\bibitem[{{Timmes} {et~al.}(1996){Timmes}, {Woosley}, \& {Weaver}}]{Tim96}
{Timmes}, F.~X., {Woosley}, S.~E., \& {Weaver}, T.~A. 1996, \apj, 457, 834

\bibitem[{{van der Marel}(2003)}]{Van03}
{van der Marel}, R.~P. 2003, in American Institute of Physics Conference
  Series, Vol. 686, The Astrophysics of Gravitational Wave Sources, ed. J.~M.
  {Centrella}, 115--124

\bibitem[{{Vink} {et~al.}(2001){Vink}, {de Koter}, \& {Lamers}}]{Vin01}
{Vink}, J.~S., {de Koter}, A., \& {Lamers}, H.~J.~G.~L.~M. 2001, \aap, 369, 574

\bibitem[{{Zampieri} {et~al.}(2004)}]{Zam04}
{Zampieri}, L., {Mucciarelli}, P., {Falomo}, R., {Kaaret}, P., {Di Stefano},
  R., {Turolla}, R., {Chieregato}, M., \& {Treves}, A. 2004, \apj, 603, 523

\bibitem[{{Zampieri} \& {Roberts}(2009)}]{Zam09}
{Zampieri}, L. \& {Roberts}, T.~P. 2009, \mnras, 400, 677

\bibitem[{{Zezas} \& {Fabbiano}(2002)}]{Zezlum}
{Zezas}, A. \& {Fabbiano}, G. 2002, \apj, 577, 726

\bibitem[{{Zezas} {et~al.}(2002)}]{Zez02}
{Zezas}, A., {Fabbiano}, G., {Rots}, A.~H., \& {Murray}, S.~S. 2002, \apj, 577,
  710

\end{thebibliography}

\end{document}